\documentclass[preprintnumbers,amsmath,amssymb,10pt]{revtex4}

\usepackage{graphicx}
\usepackage{color}
\usepackage{float}
\usepackage{amsmath}
\usepackage{appendix}
\usepackage{subfigure}

\begin{document}

\title{Superfluidity of indirect momentum space dark dipolar excitons in a double layer {with} massive anisotropic tilted semi-Dirac bands}
\author{A.\hspace{0.09cm}Nafis  Arafat$^{1,2}$, Oleg L. Berman$^{1,3}$ and Godfrey Gumbs$^{1,2,4}$}

\affiliation{$^{1}$The Graduate School and University Center, The
City University of New York, \\
365 Fifth Avenue,  New York, NY 10016, USA\\
$^{2}$Department of Physics and Astronomy, Hunter College of The City University of New York, 695 Park Avenue, \\
$^{3}$Physics Department, New York City College
of Technology,          The City University of New York, \\
300 Jay Street,  Brooklyn, NY 11201, USA \\
New York, NY 10065, USA\\
$^{4}$Donostia International Physics Center (DIPC), P de Manuel Lardizabal, 4, 20018 San Sebastian, Basque Country, Spain }

\date{\today}

\begin{abstract}
\noindent { We have theoretically investigated the spin-
and valley-dependent superfluidity properties of indirect momentum
space dark dipolar excitons in  double layers {with} massive anisotropic
tilted semi-Dirac bands in the presence of circularly polarized
irradiation. An external vertical electric field is also applied to
the structure and is responsible for tilting and gap opening for the
band structure. For our calculations we used the parameters of a
double layer of 1T$^\prime$-MoS$_2$.} Closed form analytical
expressions are presented for the energy spectrum for excitons,
their associated wave functions  and binding energies. Additionally,
we examine the effects which  the intensity and frequency  of
circularly polarized irradiation has for 1T$^\prime$-MoS$_2$     on
the effective mass of the excitons since it has been demonstrated
that the  application of an external high-frequency dressing field
tailors the crucial electronic including  the exciton binding
energy,  as well as the critical temperature for superfluidity.  We
also calculate the sound velocity in the anisotropic {weakly-interacting} Bose gas of two-component indirect momentum
space dark excitons for a double layer of 1T$^\prime$-MoS$_2$.  We
show that the critical velocity of superfluidity, the spectrum of
collective excitations, concentrations of the superfluid and normal
component, and mean field critical temperature for superfluidity are
anisotropic and {formed by} a two-component system. The
critical temperature for superfluidity is increased when the exciton
concentration and interlayer   separation are increased.  We propose
the use of phonon-assisted photoluminescence to experimentally
confirm directional superfluidity of indirect momentum space dark
excitons {in a double layer {with} massive anisotropic
tilted semi-Dirac bands.}
\end{abstract}

 \maketitle

\medskip
\par

\section{Introduction}
\label{sec1}

{Transition metal dichalcogenides (TMDCs) have been
gaining much attention in recent times for their remarkable
optoelectronic as well as transport properties.}  TMDCs have been
found to have applications for a range of fundamental phenomena
\cite{optoElectronics}. These materials can can sustain the
formation of polaritons at room temperature due to their relatively
larger binding energy \cite{starkEffect}.
Indirect momentum space dark excitons in a monolayer with
massive anisotropic tilted semi-Dirac bands have been the subject of recent experimental studies \cite{GrossPoster,OpticalReadout,StrainExciton}. So far, {massive anisotropic tilted Dirac systems such as 1T$^\prime$-MoS$_2$
are one of the more interesting 2D Dirac materials, due to their
thermodynamic stability and their ease of synthesis in the
semiconducting  phase. 1T$^\prime$-MoS$_2$} has been predicted to
display a strong quantum-spin Hall effect \cite{TiltedCones}. When a
uniform perpendicular external electric field is applied, it
exhibits valley-spin polarized Dirac bands \cite{BGR} as well as a phase
transition between topological insulator and regular band insulator
\cite{aniso1T}. The anisotropy of the energy band structure is
consequential due  to its tilted and shifted valence and conduction
Dirac bands \cite{TMDCsAbstract}.

\medskip
\par
Bose-Einstein condensation (BEC) occurs when bosons at low
temperatures occupy the lowest energy quantum state. Due to the
relatively large exciton binding energies in 2D semiconductors, such
as 1T$^\prime$-MoS$_2$, BEC and superfluidity of dipolar excitons in
double layer 1T$^\prime$-MoS$_2$  are possible. Since the de Broglie
wavelength for a 2D system is inversely proportional to the square
root of the mass of a particle, BEC is more likely to exist for
bosons of larger mass at higher temperatures than for bosons with
smaller mass. A BEC of weakly interacting particles was famously
achieved experimentally in dilute gas of alkali atoms, albeit at
challenglingly low temperatures in the nano Kelvin scale
regime~\cite{Nobel}. Therefore, BEC exists at much higher
temperatures in a Bose gas consisting of particles whose mass is
smaller than those in a system of relatively heavy alkali atoms.
{The BEC and superfluidity of dipolar bright excitons
was predicted for double layers in Ref. \cite{LYGroundState} and discussed for the double
layers of TMDCs in Refs.~\cite{Butov,HighTempSuperOleg,SuperfluidDipolar}.}

 The BEC and superfluidity of dipolar indirect momentum space excitons are intriguing
due to the fact that we expect that such excitons will be characterized by the higher lifetime than regular dipolar excitons, since their recombination with the photon emission is forbidden by selection rules \cite{Lifetime}.

\medskip
\par
In this paper, we develop an approach to study the superfluidity of
a two-component dilute Bose gas of dipolar excitons in a double
layer of massive anisotropic tilted Dirac systems. {While we perform our calculations for the specific case the double
layer of tilted 1T$^\prime$-MoS$_2$, our} approach can be applied to
other massive anisotropic tilted Dirac systems without loss of
generality. We investigate three different types of excitons. These
are based on their spin and valley polarizations as well as the way
in which they alter the effective masses and center-of-masses of the
excitons. We also study the effect which chosen parameters have on
the binding energy of the three different types of excitons.
Specifically, we analyze the way in which varying the relative value
of the perpendicular electric field affects the binding energy. From
this, we gain insight regarding the use of a number of dielectric
layers (in our case hexagonal boron nitride, h-bN), for experimental
exploration and their associated effects \cite{Rubio}.

\medskip
\par
The rest of this paper is organized as follows. In Sec.\  \ref{sec2}, the two-body problem for an electron and a hole, spatially separated in two parallel monolayers of 1T$^\prime$-MoS$_2$, is formulated. The wave function and binding energy for a single dipolar exciton in the 1T$^\prime$-MoS$_2$ double layer are calculated. Section\  \ref{sec3} is devoted to studying  whether the electron-photon dressed states by circularly polarized light, predicted in Ref. \cite{Floquet}, allow for sufficient modification of the binding energy by changing the irradiation intensity and frequency of the dressing field. In Sec.\ \ref{sec4}, we study the formation of collective excitations for spatially separated electrons and holes. In Sec.\  \ref{sec5},  we turn our attention to the  case of two-component superfluidity  exhibiting directionality induced by anisotropy of the system, for the soundlike spectrum of collective excitations, and the dependence of the sound velocity and critical temperature on the angle $\Theta$, a parameter which transforms the center-of-mass momentum into polar coordinates. Section\  \ref{sec6} discusses our results obtained for   or the superfluidity.  We propose a way to   experimentally observe the indirect-momentum space dark excitons and related phenomena. Section\  \ref{sec7} contains our concluding remarks.

\medskip
\par

\section{The interacting electron-hole pair}
\label{sec2}

We begin by computing the wavefunctions and energy eigenvalues of the massive anisotropic tilted Dirac systems by examining the general Hamiltonian \cite{GenHamiltonian}. By examining the energy dispersion relation, and modifying it to the form of the model Hamiltonian for $\mathrm{1T'-MoS_{2}}$ with tilted Dirac bands, we are able to solve exactly for the analytical solution of the system. The model  Hamiltonian within the effective mass approximation for a
single electron-hole pair in a  $\mathrm{1T'-MoS_{2}}$ double layer
is given by \cite{BGR}

\begin{eqnarray}
\hat{H}_{0} = -
\frac{\hbar^{2}}{2m_{x}^{e}}\frac{\partial^{2}}{\partial x_{1}^{2}}
- \frac{\hbar^{2}}{2m_{y}^{e}}\left[\frac{\partial}{\partial y_{1}}
- p_{e}^{(0)}\right]^{2} -
\frac{\hbar^{2}}{2m_{x}^{h}}\frac{\partial^{2}}{\partial x_{2}^{2}}
- \frac{\hbar^{2}}{2m_{y}^{h}}\left[\frac{\partial}{\partial y_{2}}
- p_{h}^{(0)}\right]^{2} + V\left(\sqrt{r^{2}+D^{2}}\right)\ ,
\label{H0}
\end{eqnarray}
where $V\left(\sqrt{r^{2}+D^{2}}\right)$ is the potential energy for electron-hole pair attraction, when the electron and hole are located in two parallel monolayers separated by distance $D$. The Hamiltonian reflects the anisotropy of the system. In Eq.~(\ref{H0}), $m_{x}^{e}$, $m_{y}^{e}$, $m_{x}^{h}$, $m_{y}^{h}$, $p_{e}^{(0)}$, and $p_{h}^{(0)}$ are effective constants which can be obtained from Eq.~(\ref{longwave}) using the method of completing squares as seen below. The constants $m_{x}^{e}$,
$m_{y}^{e}$, $m_{x}^{h}$, and $m_{y}^{h}$ are the effective masses of the electrons and holes in $x$ and $y$. The constants $p_{e}^{(0)}$, and
$p_{h}^{(0)}$ are effective constants that reflect the anisotropy of the band structure.

\begin{equation}
E(k) = E_{0} + \frac{\hbar^2}{2 m^{*}_{x}} (k_x - k_{0,x})^2 + \frac{\hbar^2}{2 m^{*}_{y}} (k_y - k_{0,y})^2
\label{EK}
\end{equation}

For the energy spectrum of Eq.(\ref{EK}) the method of completing the square is outlined below

\begin{eqnarray}
\epsilon_{\xi,s}^{\lambda}({\bf k})=  \xi |\lambda - s \alpha| \Delta +
  \left(\dfrac{\xi v_{+}^2 \hbar^2 }{2\Delta|\lambda-\alpha s|}\right)\left({k_y}^2
  +\frac{\left[  -\lambda \hbar v_{-} +\xi \hbar v_{2}\  \text{sgn} (\lambda-s\alpha) \right]({2\Delta|\lambda-\alpha s|})}{{\xi v_{+}^2 \hbar^2 }}k_{y}\right) + \left(\dfrac{\xi v_1^2   \hbar^2 }{2\Delta|\lambda-\alpha s|}\right)\left(k_{x}^2\right)
\label{longwave3}
\end{eqnarray}

We visualise the energy bands as a function of $k_x$ and $k_y$ for the different excitons outlined thoroughly in Section \ref{sec2}.B, which differ by the choice of spin and valley.

\begin{figure}[H]
\centering \subfigure(a){
\includegraphics[width=0.45\textwidth]{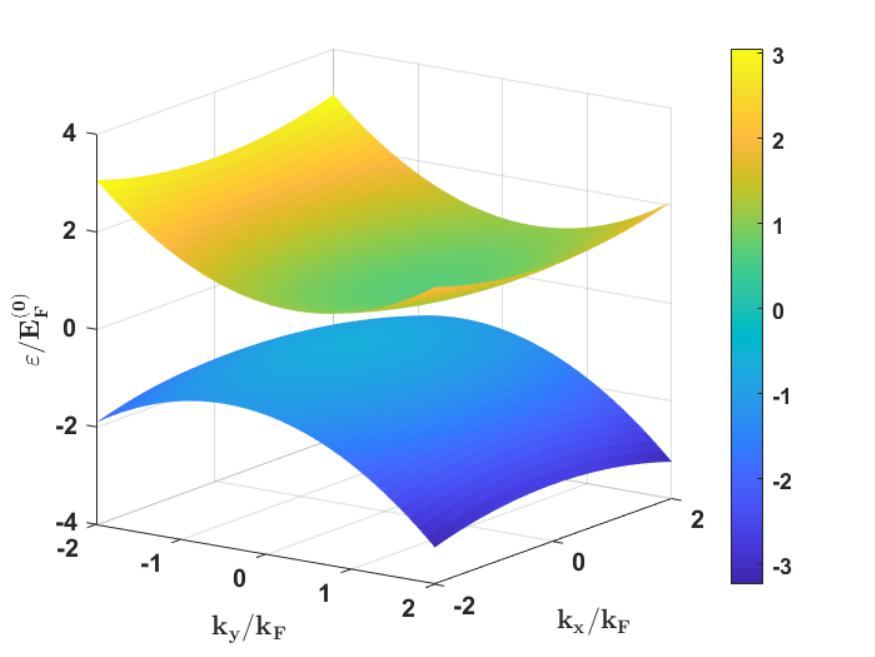}
} \subfigure(b){
\includegraphics[width=0.45\textwidth]{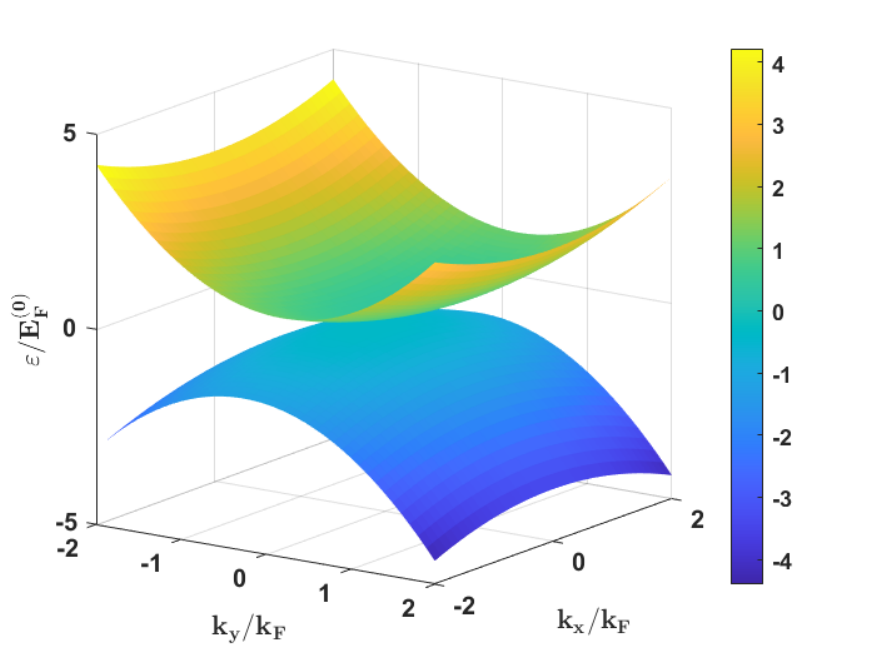}
}
\subfigure(c){
\includegraphics[width=0.45\textwidth]{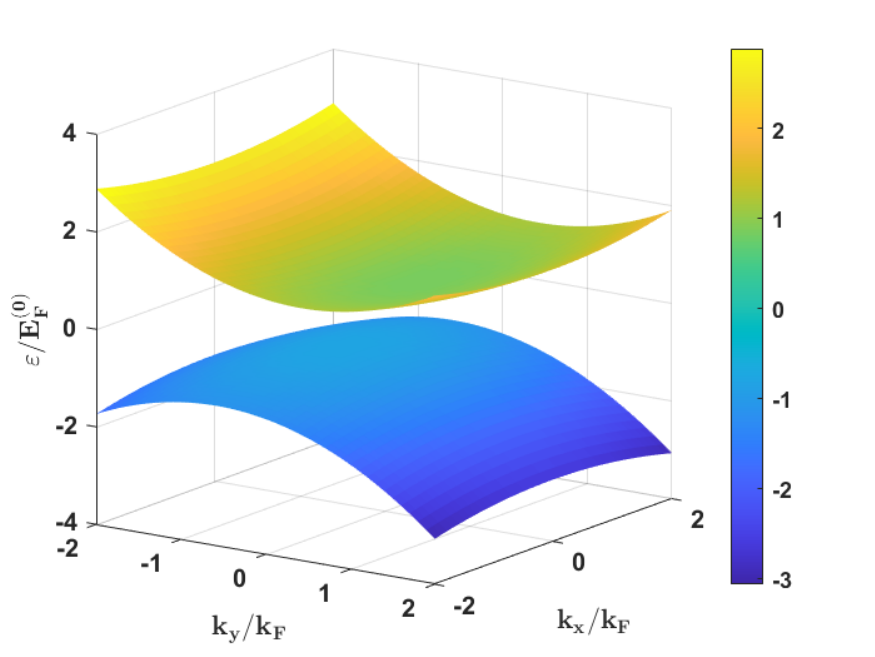}
}
\subfigure(d){
\includegraphics[width=0.45\textwidth]{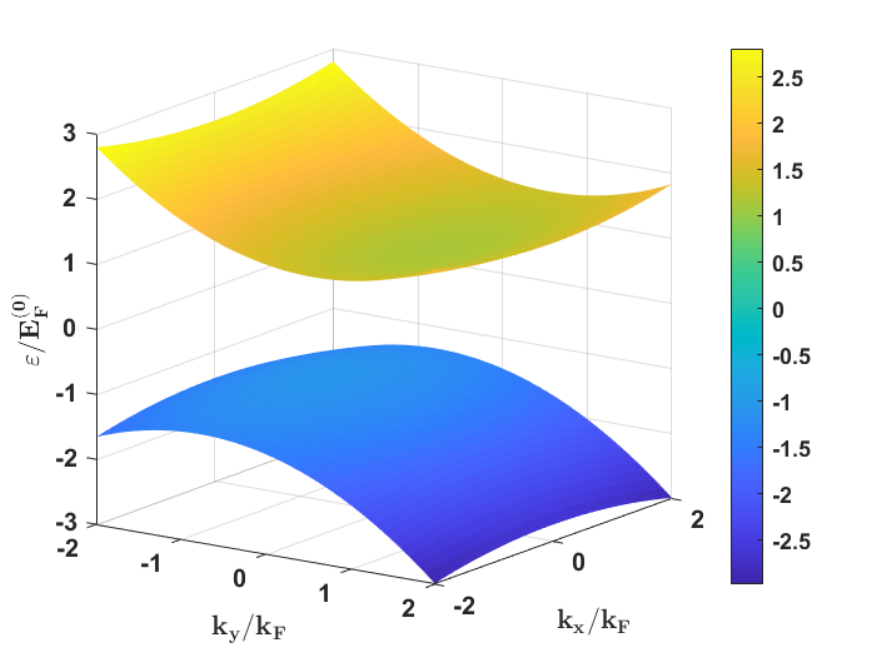}
}
\subfigure(e){
\includegraphics[width=0.45\textwidth]{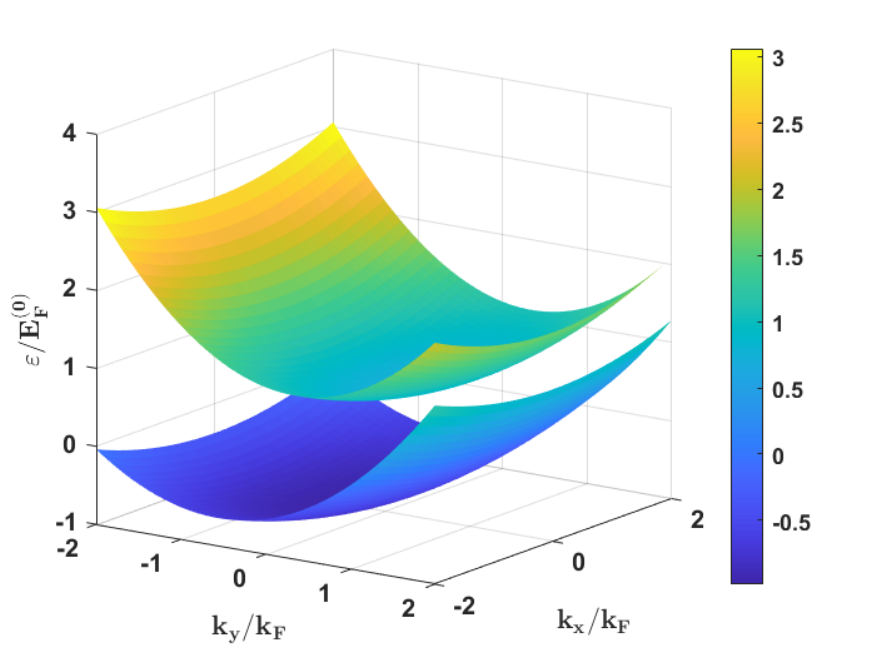}
}
\subfigure(f){
\includegraphics[width=0.45\textwidth]{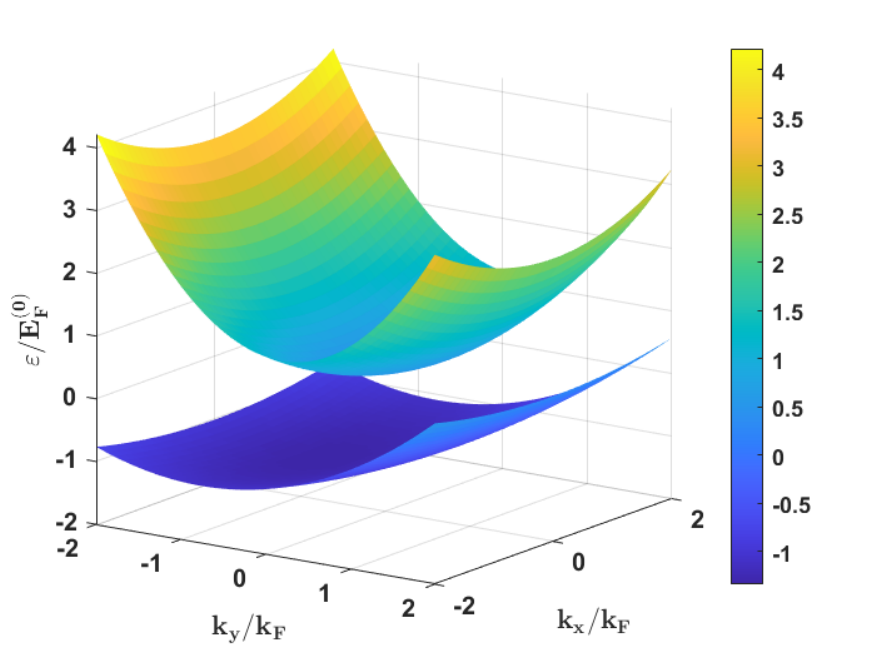}
}
\caption{(Color online) Visualization of the energy dispersion bands {for the charge carriers}, forming the different types of excitons explored in this paper, with different values of $\alpha$  - (a) Type A excitons with $\alpha = 0.1$, (b) Type A excitons with $\alpha = 0.5$, (c) Type B excitons with $\alpha = 0.1$, (d) Type B excitons with $\alpha = 0.5$, (e) Type C excitons with $\alpha = 0.1$, and (f) Type C excitons with $\alpha = 0.5$.}
\label{bands1}
\end{figure}

The bands do not form full Dirac cones but rather semi-Dirac {cones. Visualisation of the energy dispersion bands for the charge carriers, forming each of the different types of excitons is presented in Fig.\ref{bands1}}  The difference in band curvature at the origin between different types of excitons leads to the fact that different types of excitons are characterized by different effective masses. The curvature also changes for different values of $\alpha$. {We can find such value of $\alpha$, which will correspond to the higher exciton binding energy implying more stable excitons.}  Fig.\ref{bands1} also demonstrates the anisotropy of our system, which forms the basis for the directional {indirect momentum space dark exciton superfluidity presented} in Secs. \ref{sec6}-\ref{sec7}.

From Eq. (\ref{longwave3}) we can infer the expressions for the effective constants.

\begin{eqnarray}
m_{x}^{e/h} = \dfrac{\Delta|\lambda-\alpha s|}{\xi v_1^2} , \hspace{0.5cm} m_{y}^{e/h} = \dfrac{\Delta|\lambda-\alpha s|}{\xi v_{+}^2}
\label{massesXYEH}
\end{eqnarray}

\begin{eqnarray}
p_{e/h}^{(0)} =  \frac{- {i} \left[  -\lambda \hbar v_{-} +\xi \hbar v_{2}\  \text{sgn}(\lambda-s\alpha) \right]({\Delta|\lambda-\alpha s|})}{{ \xi v_{+}^2 \hbar^2 }}, \hspace{0.5cm}
\tilde{p}_{e/h}^{(0)} =  \frac{\left[  -\lambda \hbar v_{-} +\xi \hbar v_{2}\  \text{sgn}(\lambda-s\alpha) \right]({\Delta|\lambda-\alpha s|})}{{ \xi v_{+}^2 \hbar^2 }}
\label{correctionsXYEH}
\end{eqnarray}

It is worth noting that $ p_{e/h}^{(0)} $ is a  complex number, and as such the real valued constant $\tilde{p}_{e/h}^{(0)} $ was introduced, where $p_{e/h}^{(0)} = - {i} \hspace{0.05cm} \tilde{p}_{e/h}^{(0)}$, to easily identify complex numbers in differential equations. An extra energy term arises as a constant from Eq.(\ref{longwave}). Following the procedure~\cite{LandauStat,Prada}
for the separation of the relative motion of the electron-hole pair
from their center-of-mass motion one can introduce variables for the  center-of-mass of an
electron-hole pair $\mathbf{R} =(X,Y)$ and the relative motion of an electron and a hole $%
\mathbf{r} = (x,y)$, $X =  (m_{x}^{e} x_{1} + m_{x}^{h} x_{2})/(
m_{x}^{e}+ m_{x}^{h})$, \
 $Y = (m_{y}^{e} y_{1} + m_{y}^{h} y_{2})/(
m_{y}^{e}+ m_{y}^{h})$,  \   $x = x_{1} - x_{2} \ ,
 y = y_{1} - y_{2}$~,~and~$r^{2}=x^2+y^2$.
\medskip
\par
\noindent
The Schr\"{o}dinger equation with
Hamiltonian~(\ref{H0})
 has the form:
$\hat{H}_{0} \Psi(\mathbf{r}_{1},\mathbf{r}_{2}) = E\Psi(\mathbf{r}_{1},%
\mathbf{r}_{2})$, where $\Psi(\mathbf{r}_{1},\mathbf{r}_{2})$ and
$E$ are the eigenfunction and eigenenergy. One can write $\Psi(\mathbf{r}_{1},%
\mathbf{r}_{2})$ in the form $\Psi(\mathbf{r}_{1},\mathbf{r}_{2}) = \Psi(%
\mathbf{R},\mathbf{r}) = \varphi({R}) \hspace{0.1cm} \varphi(\mathbf{%
r})$, where $\varphi(\mathbf{R})$ is the wave function for the center-of-mass
and $\varphi(\mathbf{r})$
is the wave function for the electron-hole pair.
\medskip
\par
\noindent
The wave function for the center of mass is given by the 2D Schr\"{o}dinger  equation:

\begin{eqnarray}
\left[-\frac{\hbar^2}{2M_x}\frac{\partial^2}{\partial X^2_{}} - \frac{\hbar^2}{2 M_y}\frac{\partial^2}{\partial Y^2_{}} - {i} \frac{\hbar^2}{M_y}(\tilde{p}_e^{(0)} + \tilde{p}_h^{(0)})  \frac{\partial}{\partial Y_{}} + E_{0}^{(e)}+ E_{0}^{(h)} + \eta \right] \varphi (X,Y) = \varepsilon_{COM} \hspace{0.1cm} \varphi(X,Y),
\label{schrodcom2D}
\end{eqnarray}
\noindent
In Eq.(\ref{schrod2D}), $m_{x}^{e}$,
$m_{y}^{e}$, $m_{x}^{h}$, $m_{y}^{h}$, $p_{e}^{(0)}$, and $p_{h}^{(0)}$ are defined in Eqs.(\ref{massesXYEH})-(\ref{correctionsXYEH}) and the reduced masses are defined as  follows

\begin{eqnarray}
\mu_x = \frac{m^e_x m^h_x}{m^e_x + m^h_x}, \hspace{0.6cm}
\mu_y = \frac{m^e_y m^h_y}{m^e_y + m^h_y}
\label{effectiveMassesXY}
\end{eqnarray}

\begin{eqnarray}
\mu^{*}_y = \frac{m^e_y m^h_y}{m^e_y p^{(0)}_h + m^h_{y} p^{(0)}_e}, \hspace{0.6cm}
\tilde{\mu}^{*}_y = \frac{m^e_y m^h_y}{m^e_y \tilde{p}^{(0)}_h + m^h_{y} \tilde{p}^{(0)}_e}
\end{eqnarray}

\begin{eqnarray}
\eta =  \frac{\hbar^{2}}{2}\frac{m^{e}_y {{\tilde{p}^{(0)}_h}{}^2} + m^h_y {{\tilde{p}^{(0)}_e}}{}^{2}}{m^{e}_y m^{h}_y}
\label{reducedmass}
\end{eqnarray}

It is worth noting that $\eta$ is a positive, real number and $ {\mu}^{*}_y  $ is a complex number. Consequently,   we introduce the real valued constant $\tilde{\mu}^{*}_y$ to easily identify complex numbers in equations. The relationship between them is $ {\mu}^{*}_y = {i} \hspace{0.05cm}\tilde{\mu}^{*}_y $. The wave function for the relative motion of the electron-hole pair is given by the 2D Schr\"{o}dinger  equation:

\begin{eqnarray}
\left[-\frac{\hbar^2}{2\mu_x}\frac{\partial^2}{\partial x^2_{}} - \frac{\hbar^2}{2\mu_y}\frac{\partial^2}{\partial y^2_{}} + \frac{\hbar^2}{\mu^*_y}\frac{\partial}{\partial y_{}} + V\left(\sqrt{r^2 + D^2}\right)\right] \varphi (x,y) = \varepsilon_{rel} \hspace{0.1cm} \varphi(x,y),
\label{schrod2D}
\end{eqnarray}

\begin{eqnarray}
\left[-\frac{\hbar^2}{2\mu_x}\frac{\partial^2}{\partial x^2_{}} - \frac{\hbar^2}{2\mu_y}\frac{\partial^2}{\partial y^2_{}}  - {i}\frac{\hbar^2}{\tilde{\mu}^*_y}\frac{\partial}{\partial y_{}} + V\left(\sqrt{r^2 + D^2}\right)\right] \varphi (x,y) = \varepsilon_{rel} \hspace{0.1cm} \varphi(x,y),
\label{schrod2Dreal}
\end{eqnarray}


\subsection{Wave function and binding energy of an exciton}
\medskip
\par

{Electron-hole interaction in a double layer is
discussed in Appendix~B.} Substituting Eq.\ (\ref{expand}) with
parameters in Eq.\ (\ref{V0g}) for {Coulomb potential,}
and using $ r^2 = x^2 + y^2 $, one obtains an equation which has the
form of the Schr\"{o}dinger equation for a 2D anisotropic harmonic
oscillator. This allows us to carry out separation of variables to
obtain two independent equations

\begin{eqnarray}
- \frac{\hbar^2}{2 \mu_x} \frac{\partial^2}{\partial x^2} \psi (x) + \gamma x^2 \psi(x) = \left( \varepsilon_x + \frac{V_0}{2}\right)\psi(x)
\label{schx1D}
\end{eqnarray}

\begin{eqnarray}
- \frac{\hbar^2}{2 \mu_y} \frac{\partial^2}{\partial y^2} \psi (y) + \frac{\hbar^2}{\mu^{*}_y}\frac{\partial}{\partial y} \psi (y) + \gamma y^2 \psi(y) = \left( \varepsilon_y + \frac{V_0}{2}\right)\psi(y)
\label{schy1D}
\end{eqnarray}
\noindent
The wavefunction and eigenenergies of Eq.(\ref{schx1D}) are obtained simply since it is the well-studied solution to a 1D harmonic oscillator.
The process for solving Eq.(\ref{schy1D}) is the following. First we assume that $ \psi(y)$ has the form of two functions multiplied.
\begin{eqnarray}
\psi(y) = f(y) \cdot g(y)
\label{schy1Dcomp}
\end{eqnarray}
\noindent
\noindent
allowing us to rewrite Eq.(\ref{schy1D}) into the following form

\begin{eqnarray}
a_2 \hspace{0.05cm} g(y) \frac{\partial^2}{\partial y^2} f (y) + \left(2 a_2 \frac{\partial}{\partial y} g(y) +  a_1 g(y) \right)\frac{\partial}{\partial y} f (y) + \left(a_2 \frac{\partial^2}{\partial y^2} g(y) + a_1 \frac{\partial}{\partial y} g(y)  + (\gamma y^2-E_y) g(y) \right) f(y) = 0 \  ,
\label{schy1DSasha}
\end{eqnarray}
\noindent
where

\begin{eqnarray}
E_y = \varepsilon_y + \frac{V_0}{2}, \hspace{0.2cm} a_2 = -\frac{\hbar^2}{2 \mu_y}, \hspace{0.2cm} a_1 = \frac{\hbar^2}{\mu_y^*} =  - {i} \hspace{0.02cm} \tilde{a}_1, \hspace{0.2cm} \tilde{a}_1 = \hspace{0.04cm} \frac{\hbar^2}{\tilde{\mu}_y^*}
\end{eqnarray}

\noindent
By imposing the condition that the coefficient of the first derivative of $f(y)$ in Eq.(\ref{schy1DSasha}) must be equal to zero, we find the form of $g(y)$.

\begin{eqnarray}
2 a_2 \frac{\partial}{\partial y} g(y) - {i} \hspace{0.04cm}\tilde{a}_1 g(y) = 0, \hspace{0.2cm} g(y) = F e^{{i} \hspace{0.03cm}\frac{\tilde{a}_1 y}{2a_2}}  \  ,
\label{gy}
\end{eqnarray}
\noindent
where $F $ is an undetermined constant. Inserting our result from Eq.(\ref{gy}) into Eq.(\ref{schy1DSasha}) we obtain an equation that determines $f(y)$. This equation is of the form of a 1D harmonic oscillator.

\begin{eqnarray}
a_2 \frac{\partial^2}{\partial{y}^2}f(y) + \gamma y^2 f(y) = \left(E_y - \frac{\tilde{a}_1^2}{4a_2}  \right) f(y)
\label{fy}
\end{eqnarray}

\noindent
Using the condition of normalization
\(
\int_{-\infty}^{\infty} |\psi(y)|^2 \,dy = 1
\)
 we find $F$ = 1.
The normalized eigenfunctions for Eq.  (\ref{schx1D}) and Eq.\   (\ref{schy1D}) are given by

\begin{eqnarray}
\psi_n (x) = \frac{1}{\pi^{1/4} a_x^{1/2}} \frac{1}{\sqrt{2^n n!}} e^{-x^2/(2a_x^2)} H_n \left(\frac{x}{a_x}\right)
\label{schx1Dsol}
\end{eqnarray}

\begin{eqnarray}
\psi_m (y) = e^{-{i}\frac{\mu_y y}{\tilde{\mu}_y^{*}}} \frac{1}{\pi^{1/4} a_y^{1/2}} \frac{1}{\sqrt{2^m m!}} e^{-y^2/(2a_y^2)} H_m \left(\frac{y}{a_y}\right)
\label{schy1Dsol}
\end{eqnarray}

\noindent
where $ n = 0, 1, 2, 3...$ and $ m = 0, 1, 2, 3...$ are the quantum numbers, $ H_n (\frac{x}{a_x}) $, and $ H_m (\frac{y}{a_y}) $ are Hermite polynomials, and  $ a_x = (\hbar/(\sqrt{2 \mu_x \gamma})^{1/2}$ and $ a_y = (\hbar/(\sqrt{2 \mu_y \gamma})^{1/2}$.
\noindent
The overarching wavefunction for the relative motion of the electron-hole pair in Eq.(\ref{schrod2D}) is thus given by
\(
\psi_{n,m} (x,y) = \psi_{n} (x) \psi_{m} (y) .
\)
The corresponding eigenenergies are

\begin{eqnarray}
\varepsilon_{n}^{(x)} = -\frac{V_0}{2} + \hbar \sqrt{\frac{2 \gamma}{\mu_x}}(n + 1/2), \hspace{0.6cm} \varepsilon_{m}^{(y)} = -\frac{V_0}{2} + \hbar \sqrt{\frac{2 \gamma}{\mu_y}}(m + 1/2) - \tilde{\chi}
\label{schx1Deigenenergy}
\end{eqnarray}

\noindent
where the constant term at the end of Eq. (\ref{schy1Deigenenergy}) is given by
\(  \tilde{\chi} = \frac{\hbar^2 \mu_y}{2{{{\tilde{\mu}_y{}^*}}}^2} .
\)
The total eigenergy of the relative motion of the electron-hole pair can be expressed as

\begin{eqnarray}
\varepsilon_{nm} = \varepsilon_{n}^{(x)}  + \varepsilon_{m}^{(y)}
\label{schy1Deigenenergy}
\end{eqnarray}
\noindent
where $ n = 0, 1, 2, 3. ...$ and $ m = 0, 1, 2, 3. ...$ \\
\medskip
\par
\noindent

The variables $X$ and $Y$ in the Schr\"{o}dinger equation for center of mass can be separated in Eq.({\ref{schrodcom2D}}) to obtain the following equations

\begin{eqnarray}
\left[-\frac{\hbar^2}{2M_x}\frac{\partial^2}{\partial X^2_{}} \right] \phi (X) = \varepsilon_{X} \hspace{0.1cm} \phi(X),
\label{schrodcom2DX}
\end{eqnarray}
{
\begin{eqnarray}
\left[- \frac{\hbar^2}{2 M_y}\frac{\partial^2}{\partial Y^2_{}} -
{i} \hspace{0.02cm} \frac{\hbar^2}{M_y}(\tilde{p}_e^{(0)} +
\tilde{p}_h^{(0)})  \ \  \frac{\partial}{\partial Y_{}} \right] \phi
(Y) = \varepsilon_{Y} \hspace{0.1cm} \phi(Y), \label{schrodcom2DY}
\end{eqnarray}
\noindent } where $M_X = m_x^e + m_x^h $, and $ M_Y = m_y^e + m_y^h
$. We will also be using the notation $\tau = \tilde{p}_e^{(0)} +
\tilde{p}_h^{(0)}$ for convenience. The equation for the center of
mass in $X$, Eq.(\ref{schrodcom2DX}), is simply that of a free
particle. The solution of
Eq.(\ref{schrodcom2DX})-(\ref{schrodcom2DY}), $ \phi (X)$ and $ \phi
(Y)$ is given by

\begin{eqnarray}
\phi(X) = \frac{1}{\sqrt{L_X}} e^{i \frac{P_X }{\hbar} X}, \hspace{0.4cm} \phi(Y) = \frac{1}{\sqrt{L_Y}} e^{i (\frac{P_Y}{\hbar} - \tau) Y}
\label{schrodcom2DXsol}
\end{eqnarray}
\noindent
The center of mass eigenenergies $ \varepsilon_X$ and $ \varepsilon_Y$ are given by

\begin{eqnarray}
\varepsilon_X = \frac{P_X^2}{2M_X}, \hspace{0.4cm} \varepsilon_Y =  \frac{P_Y^2}{2M_Y} - \frac{\hbar^2 \tau^2}{2 M_Y}
\label{schrodcom2DXEnergy}
\end{eqnarray}
\noindent
The total eigenfunction and eigenergy for the center of mass equation  are given by

\begin{eqnarray}
\phi (X,Y) = \phi (X) \hspace{0.03cm} \phi(Y), \hspace{0.4cm} \varepsilon_{COM} = \varepsilon_X + \varepsilon_Y + E_{0}^{(e)} + E_{0}^{(h)} + \eta
\label{schrodWavefcn}
\end{eqnarray}

\subsection{Types of excitons and tunability of parameters}
\medskip
\par

The location of the electron and hole in the vicinity of two inequivalent valleys $K$ and $K^\prime$ is characterized by its Dirac point ($\lambda = \pm 1$). The effects of the mass anisotropy on the exciton binding energy has been demonstrated before \cite{Rodin}. This parameter, along with the particle's spin, determines the effective masses of the electron and hole, and also the effective mass of the excitons. Using different combinations of these two parameters, we are able to categorize the excitons of three types, with three different reduced masses. The first type of exciton arises when the location of the Dirac point corresponds directly with the sign of the spin of the electron and hole.  We refer to these as {\bf Type A} excitons. The second type of exciton occurs when the location of the Dirac point is opposite the sign of the spin of the electron and hole. We term these {\bf Type B} excitons. It is important to note that the spin of the hole, when computed, is opposite that of the electron state being annihilated. These types of excitons can be formed by employing circularly polarized light. Table I below outlines the effective masses of the electron and hole along the $x$ and $y$ directions.

\medskip
\par

\begin{table}[h!]
\centering
\begin{tabular}{||c c c c c c c c||}
 \hline
 $\lambda_e$ & $\lambda_h$ & $s_e$ & $s_h$ & $m_x^e$ & $m_x^h$ & $m_y^e$ & $m_y^h$ \\ [0.5ex]
  \hline\hline
 + & + & + & \textbf{-} & $\frac{\Delta|1 - \alpha|}{\xi v_1^2}$ & $\frac{\Delta|1 - \alpha|}{\xi v_1^2}$ & $\frac{\Delta|1 - \alpha|}{\xi v_p^2}$ & $\frac{\Delta|1 - \alpha|}{\xi v_p^2}$\\

  \textbf{-} & \textbf{-} & \textbf{-} & + & $\frac{\Delta|1 - \alpha|}{\xi v_1^2}$ & $\frac{\Delta|1 - \alpha|}{\xi v_1^2}$ & $\frac{\Delta|1 - \alpha|}{\xi v_p^2}$ & $\frac{\Delta|1 - \alpha|}{\xi v_p^2}$\\ [1ex]

 + & + & \textbf{-} & + & $\frac{\Delta|1 + \alpha|}{\xi v_1^2}$ & $\frac{\Delta|1 + \alpha|}{\xi v_1^2}$ & $\frac{\Delta|1 + \alpha|}{\xi v_p^2}$ & $\frac{\Delta|1 + \alpha|}{\xi v_p^2}$\\

\textbf{-} & \textbf{-} & + & \textbf{-} & $\frac{\Delta|1 + \alpha|}{\xi v_1^2}$ & $\frac{\Delta|1 + \alpha|}{\xi v_1^2}$ & $\frac{\Delta|1 + \alpha|}{\xi v_p^2}$ &  $\frac{\Delta|1 + \alpha|}{\xi v_p^2}$\\[1ex]
\hline
\end{tabular}
\caption{Combinations of $\lambda$ and assorted spins for Type {A} and {B} excitons and their respective effective masses. The first two rows designate the properties and effective masses of electrons and holes for Type {A} excitons, and the last two rows designate the properties and effective masses of electrons and holes for  Type {B} excitons.}
\label{table:1}
\end{table}

Table II presents the combinations of excitons induced by linearly polarized light. These excitons allow for electrons and holes of different spins to exist in two different valleys. These are designated by the signs of the Dirac points. The reduced masses and center-of-mass masses of these excitons group them into one type of exciton which we shall refer to as {\bf Type C} excitons.

\begin{table}[h!]
\centering
\begin{tabular}{||c c c c c c c c||}
 \hline
 $\lambda_e$ & $\lambda_h$ & $s_e$ & $s_h$ & $m_x^e$ & $m_x^h$ & $m_y^e$ & $m_y^h$ \\ [0.5ex]
  \hline\hline
+ & \textbf{-} & + & \textbf{-} & $\frac{\Delta|1 - \alpha|}{\xi v_1^2}$ & $\frac{\Delta|1 + \alpha|}{\xi v_1^2}$ & $\frac{\Delta|1 - \alpha|}{\xi v_p^2}$ & $\frac{\Delta|1 + \alpha|}{\xi v_p^2}$\\

+ & \textbf{-} & \textbf{-} & + & $\frac{\Delta|1 + \alpha|}{\xi v_1^2}$ & $\frac{\Delta|1 - \alpha|}{\xi v_1^2}$ & $\frac{\Delta|1 + \alpha|}{\xi v_p^2}$ & $\frac{\Delta|1 - \alpha|}{\xi v_p^2}$\\

\textbf{-} & + & + & \textbf{-} & $\frac{\Delta|1 + \alpha|}{\xi v_1^2}$ & $\frac{\Delta|1 - \alpha|}{\xi v_1^2}$ & $\frac{\Delta|1 + \alpha|}{\xi v_p^2}$ &  $\frac{\Delta|1 - \alpha|}{\xi v_p^2}$\\

 \textbf{-} & + & \textbf{-} & + & $\frac{\Delta|1 - \alpha|}{\xi v_1^2}$ & $\frac{\Delta|1 + \alpha|}{\xi v_1^2}$ & $\frac{\Delta|1 - \alpha|}{\xi v_p^2}$ & $\frac{\Delta|1 + \alpha|}{\xi v_p^2}$\\ [1ex]
 \hline
\end{tabular}
\caption{Combinations of $\lambda$ and assorted spins for Type {C} excitons and their respective masses.}
\label{table:1}
\end{table}
\noindent

{It is clear} from Table II that the effective masses of
the electron and hole, in the $x$ and $y$ directions interact in
such a way that all presented combinations lead to the same
effective reduced mass and effective center of mass. These are all
Type~C excitons. Using the data outlined in Tables I and II, we are
able to then calculate the effective masses in terms of the mass of
a free electron ($m_0$), and proceed to represent them graphically
as functions of $\alpha$, seen below in Fig. 1.
\medskip
\par
\newpage

\begin{figure}[H]
\centering \subfigure(a){
\includegraphics[width=0.46\textwidth]{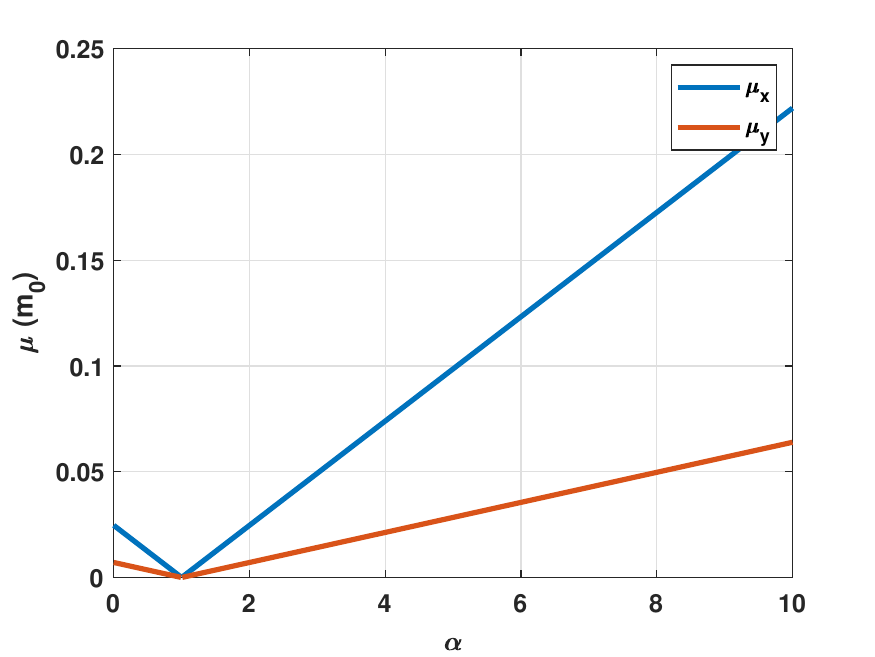}
} \subfigure(b){
\includegraphics[width=0.46\textwidth]{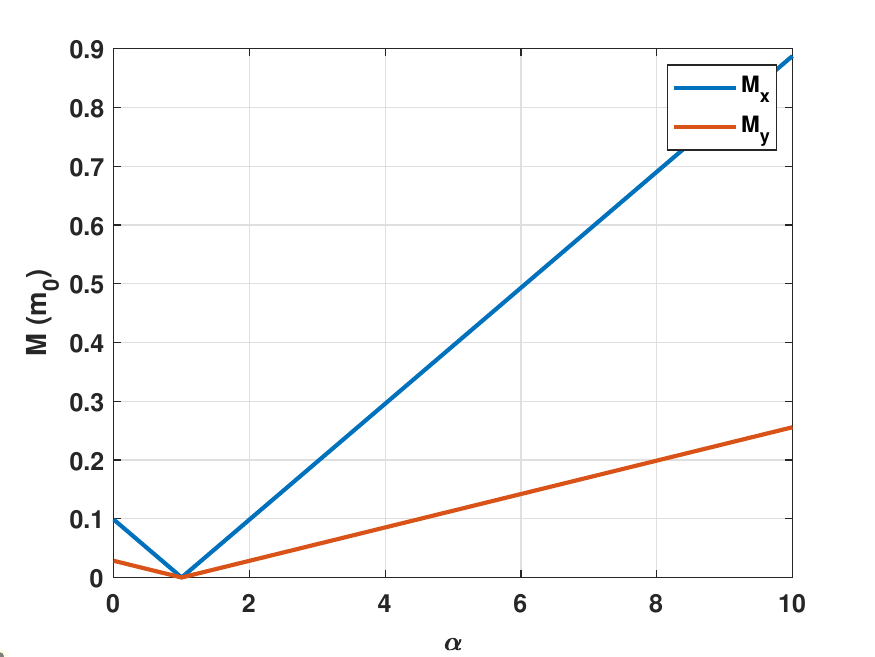}
}
\subfigure(c){
\includegraphics[width=0.46\textwidth]{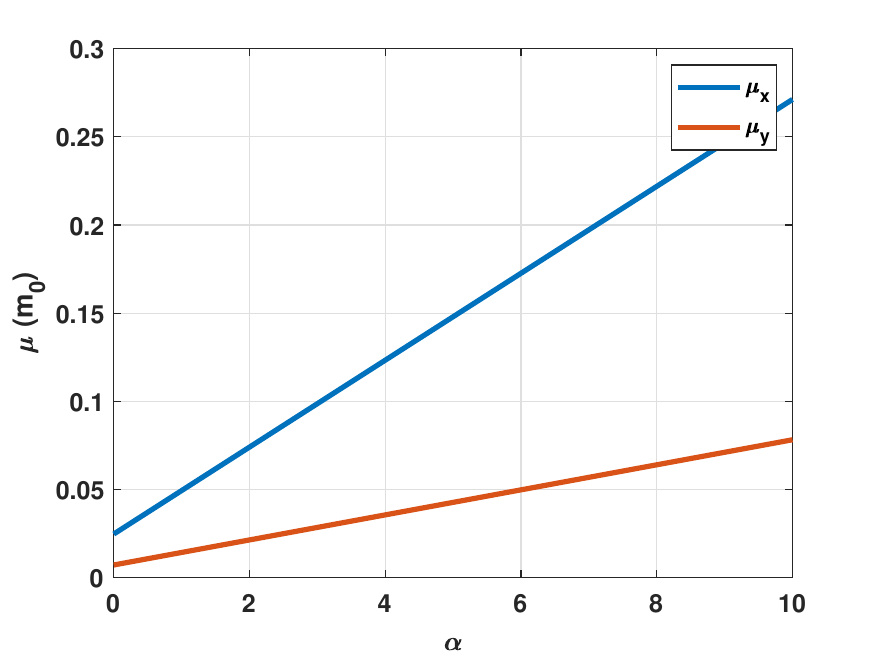}
}
\subfigure(d){
\includegraphics[width=0.46\textwidth]{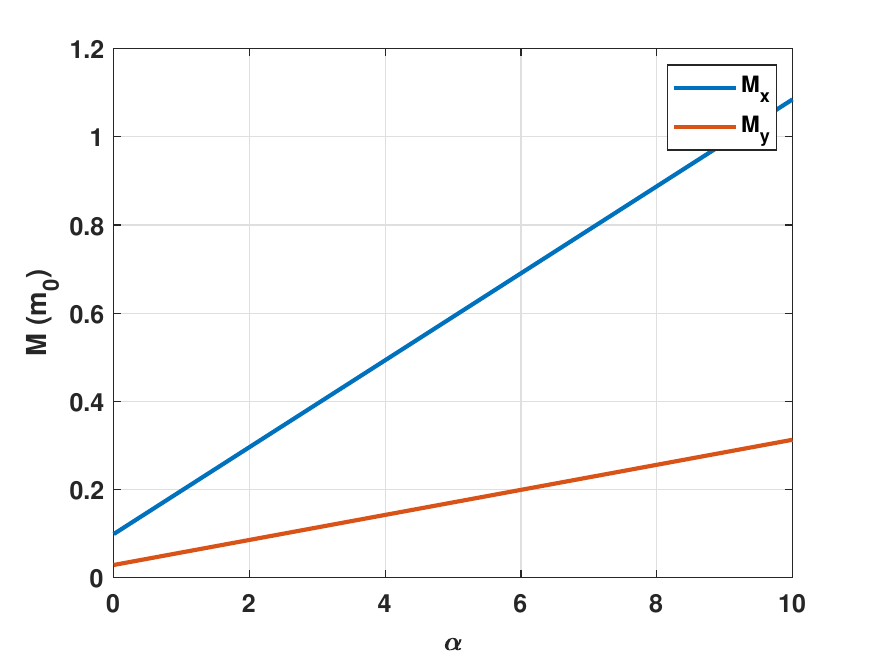}
}
\caption{(Color online) Visualization of the dependence on $\alpha$ of the (a) reduced mass of Type A excitons (b) center-of-mass of Type A excitons (c) reduced mass of Type B excitons, and (d) center-of-mass of Type B excitons. We employed the effective mass definitions of Eqn. (\ref{effectiveMassesXY}) in the generation of these figures.}
\label{Fig1}
\end{figure}

\medskip
\par
\noindent
The relationship between the reduced masses and center-of-mass masses for other excitons depends on both the chosen spin and valley. We will also see the out-of-plane electric field, $\alpha$ can be used to tune the exciton binding energies, by tuning the effective mass parameters \cite{Katsnelson, first}. Specifically, for the three types of excitons, we see different behavior. For {\bf Type A} excitons, we observe that both the reduced mass and  that for the center-of-mass decrease linearly from $ 0 \leq \alpha \leq 1.0 $ with  an increase of $\alpha$, while for $\alpha \geq 1$, the reduced mass and that for the
center-of-mass increase linearly with an increase of $\alpha$. For {\bf Type B} excitons, the reduced mass and that for the center-of-mass increase linearly for all $\alpha$, with increase of $\alpha$. Notably, for both {\bf Type A} and {\bf Type B} excitons, the reduced mass in \textbf{x} ($\mu_x$) and center-of-mass in \textbf{X} (M$_X$), are larger than the reduced mass in \textbf{y} ($\mu_y$) and center-of-mass in \textbf{Y} (M$_Y$).

\begin{figure}[h!]

\centering \subfigure(a){
\includegraphics[width=0.46\textwidth]{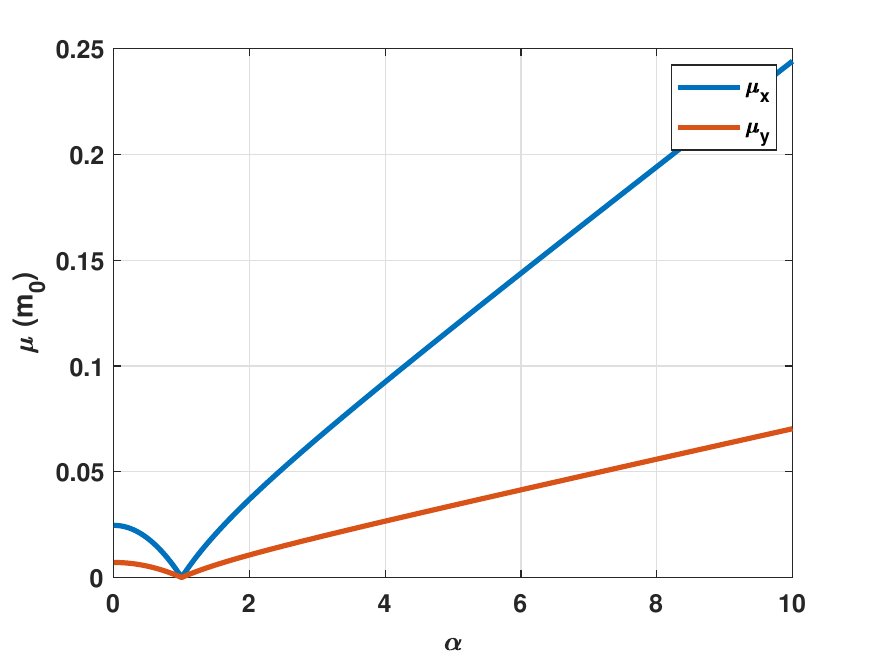}
} \subfigure(b){
\includegraphics[width=0.46\textwidth]{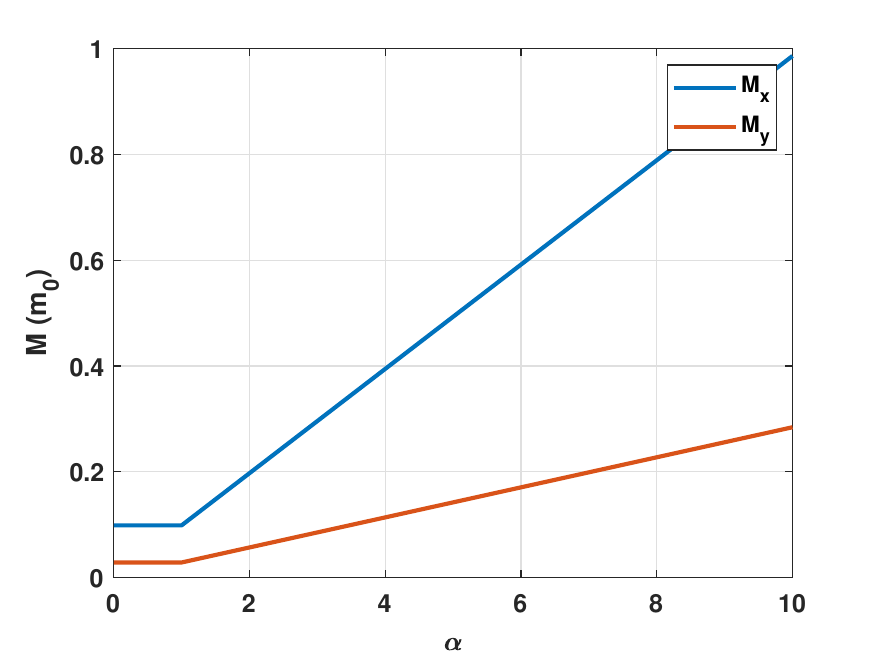}
}
\caption{(Color online) The dependence on $\alpha$ for Type {C} excitons (a) reduced exciton
mass and (b) center-of-mass exciton mass. We employed the definition of the effective mass definitions of Eqn.(\ref{effectiveMassesXY}) in the generation of these figures.}
\label{Fig1}
\end{figure}

\medskip
\par

\noindent For Type {C} excitons we observe similar
behaviour, seen above in Fig. 2. The reduced mass decreases
non-linearly from $0 \leq \alpha \leq 1$ and increases non-linearly
for $\alpha \geq 1$. The center-of-mass mass is constant for $0 \leq
\alpha \leq 1$ and increases linearly for $\alpha \geq 1$, with
increasing $\alpha$. Based on these results we can conclude that
using a large value of $\alpha$ is beneficial for our binding
energies, since it leads to larger values of binding energies, thus
increasing the stability of the excitons.
\medskip
\par
\medskip
\par

\newpage
\subsection{Binding Energy of excitons with $h$-BN dielectric }

\noindent
The binding energy corresponding to the energy spectrum of an electron and hole described by Eq. (\ref{schy1Deigenenergy}), is given by

\begin{eqnarray}
B = - \mathcal{E}_{00}= V_0 - \hbar \sqrt{\frac{\gamma}{2 \mu_x}}- \hbar \sqrt{\frac{\gamma}{2 \mu_y}} = V_0 - \hbar \sqrt{\frac{\gamma}{2 \mu_0}}
\label{bindingEnergy}
\end{eqnarray}
\noindent
In Eq.\  (\ref{bindingEnergy}), we define the quantity $ \mu_0 = \frac{\mu_x \mu_y}{(\sqrt{\mu_x} + \sqrt{\mu_y})^2}$ for convenience. In Fig.\ 3, we note that the binding energy of  {\bf Type A} excitons is decreased  for increasing $D$ with chosen $\alpha$, and is increased for increasing $\alpha$, for chosen $D$. In our structure, we have thin sheets of h-BN dielectric between layers and occupy the the interlayer region of separation~$D$. The dielectric is inserted to reduce degradation of the heterostructure and reduce the photoluminescence linewidth (PL) \cite{hBN}. Each layer of h-BN is 0.33 nm thick and as such for experimental considerations, we restrict the number of layers to ten, or fewer, so as to be experimentally viable, while also forming a bound state of the exciton. Consequently, a desirable choice of parameters for the system includes a large binding energy, comparable in magnitude to prior results \cite{SuperfluidDipolar}, and a low interlayer separation $D$ about  3.3 nm. We find that for larger values of $\alpha$, we can obtain larger binding energy for $D = 20.0$~nm which corresponds to a value of $\alpha = 6.0$. For all of {\bf Type A, B, C} excitons, the choice of $\alpha = 6.0$ or higher, allows for larger binding energy with a desirably low interlayer separation. We see that the larger the reduced mass, the greater the binding energy, as per our theoretical predictions. Too large an interlayer separation is not desirable since that would require a very large number of layers of h-BN dielectric.

\medskip
\par
For each type of our excitons  ({\bf  Type A, B, C}), we find that a larger value of $\alpha$ satisfies the criteria for a small interlayer separation $D$, with a growing value of the binding energy, which is preferable for stability of the exciton. This behavior is illustrated in Figs. 3 through 5. When $\alpha = 35.0$, we find that the binding energy of {\bf Type A, B} and {\bf C} excitons is 33.9~meV, 35.5 meV and 34.7 meV, respectively, for an interlayer separation of $D = 3.3$ nm, which corresponds to $N_L = 10$ layers of $h$-BN. It is worth noting that the corresponding value of $D_0$ for {\bf Type A, B} and {\bf C} excitons is 0.63 nm, 0.59 nm, and 0.62 nm, respectively. Consequently, the first-order Taylor  series expansion is valid for the Coulomb potential. These results are as expected based on prior results \cite{Reichman}.

\medskip
\paragraph{
}
\begin{figure}[H]

\centering
\subfigure(a){
\includegraphics[width=0.46\textwidth]{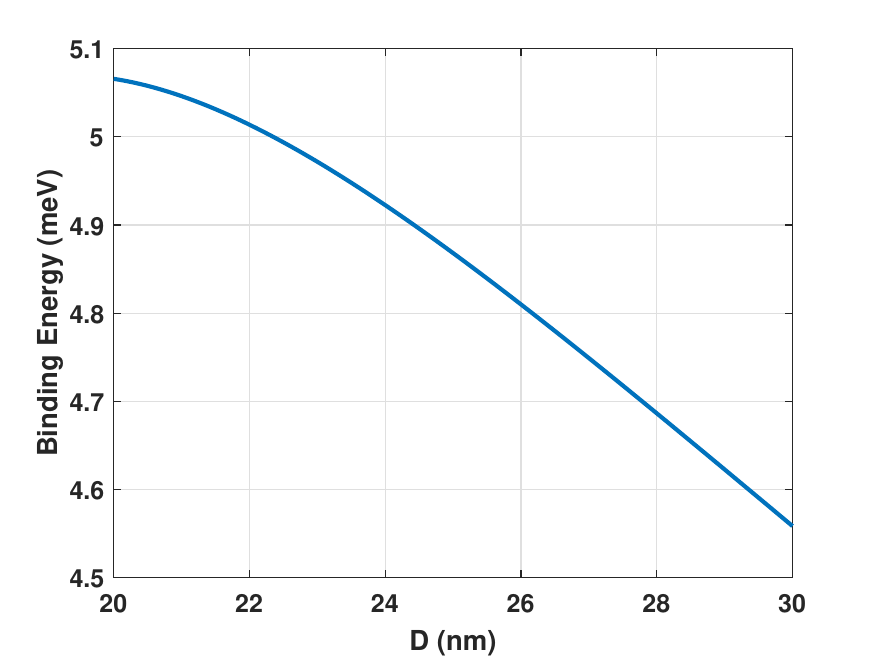}
}
\subfigure(b){
\includegraphics[width=0.46\textwidth]{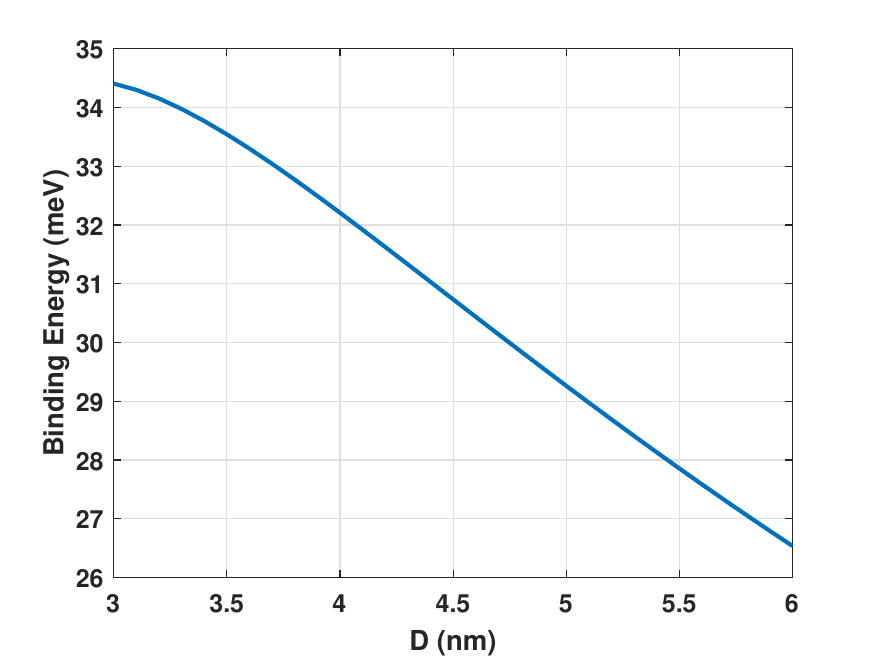}
}
\caption{The binding energy of Type \textbf{A} excitons as functions of the interlayer separation D, at different values of $\alpha$ - (a) $\alpha = 6.0 $, and (b) $\alpha = 35.0$. We employed the binding energy definition of Eqn.(\ref{bindingEnergy}), along with the effective mass definitions of Eqn. (11)-(15), in the generation of these figures.}

\label{FigA}
\end{figure}

\begin{figure}[H]

\centering
\subfigure(a){
\includegraphics[width=0.46\textwidth]{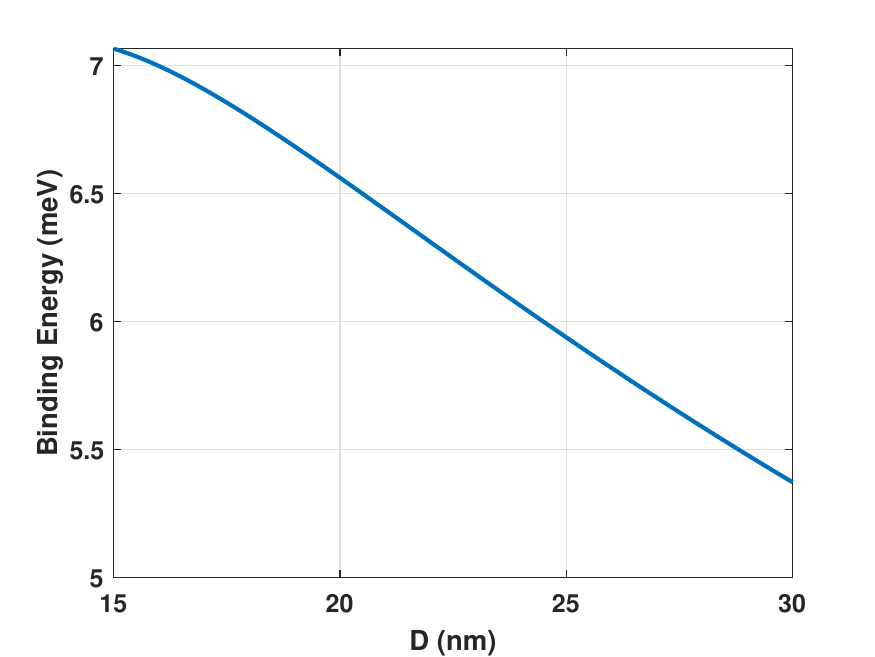}
}
\subfigure(b){
\includegraphics[width=0.46\textwidth]{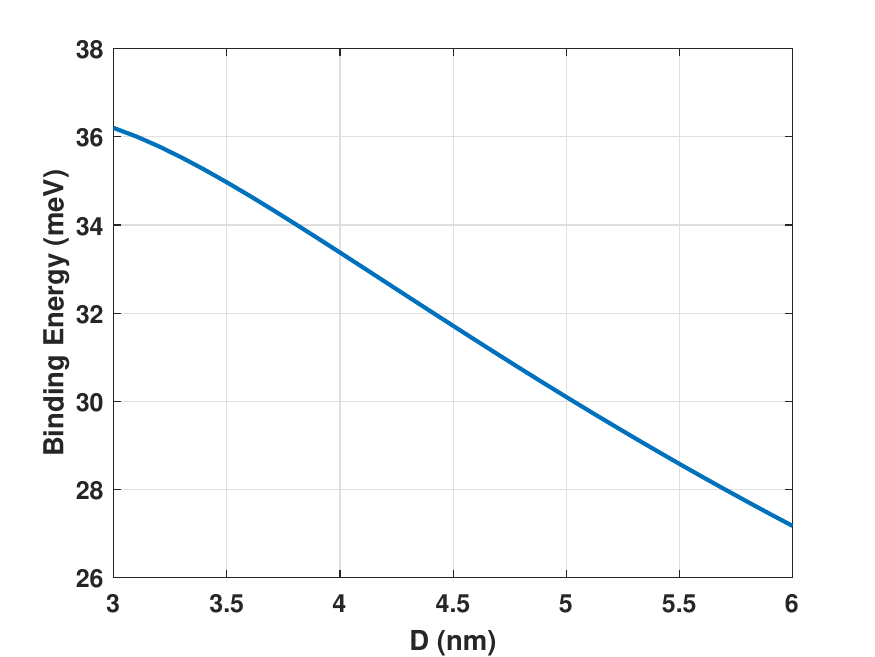}
}
\caption{The binding energy for Type {B} excitons as functions of the interlayer separation $D$, at different values of $\alpha$ - (a)~$\alpha = 6.0 $, and (b) $\alpha = 35.0$. We employed the binding energy definition of Eqn.(\ref{bindingEnergy}), along with the effective mass definitions of Eqn.(\ref{effectiveMassesXY}), in the generation of these figures.}

\label{FigB}
\end{figure}

\begin{figure}[H]
\centering
\subfigure(a){
\includegraphics[width=0.46\textwidth]{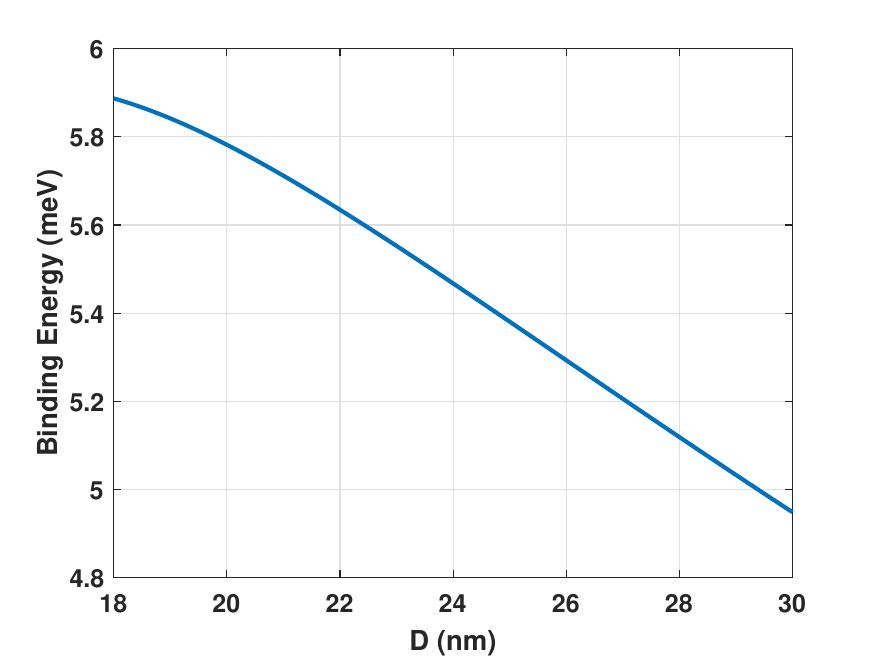}
}
\subfigure(b){
\includegraphics[width=0.46\textwidth]{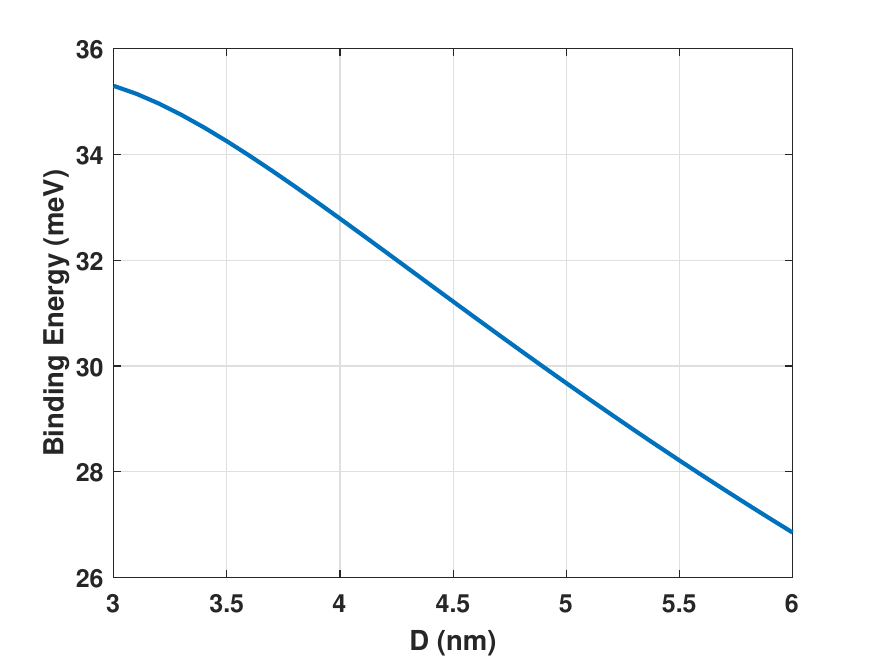}
}
\caption{The binding energy of Type \textbf{C} excitons as functions of the interlayer separation $D$, at different values of $\alpha$ - (a) $\alpha = 6.0 $, and (b) $\alpha = 35.0 $. We employed the binding energy definition of Eqn.(\ref{bindingEnergy}), along with the effective mass definitions of Eqn.(\ref{effectiveMassesXY}), in the generation of these figures.}

\label{FigC}
\end{figure}

\section{Floquet engineering of tilted and gapped Dirac band structure}
\label{sec3}

\medskip
\par

\subsection{Circularly-polarized dressing field}

It has been suggested that applying an external high-frequency optical dressing field of variable intensity, within the off-resonance regime, can modify the band structure, anisotropy, and band gaps of 1T$^\prime$-MoS$_2$~\cite{Floquet}. Additionally, it was demonstrated that the electron-photon dressed states vary strongly with the polarization of the applied irradiation and reflect a full complexity of the low-energy Hamiltonian for non-irradiated material. Employing the numerical results for the dressed states obtained by using circularly-polarized irradiation, derived in~ Ref. \cite{Floquet}, we  calculate the effective masses of the electron and hole using the following dispersion of irradiated subbands given by Eq.~(\ref{energyIrr21}), with the hope of leveraging the frequency and intensity variables that allow for higher binding energies, and parameterise the superfluidity \cite{Floquet}.

\begin{eqnarray}
\varepsilon_{\tau = \pm 1} (\textbf{k} | \xi,s) = - \hbar \xi  \upsilon_{-} k_y \pm \hbar \sqrt{ \tilde{\Delta}^2_{\xi | s}(\lambda_0) + [(\xi -s r_E) \Delta_0 + \upsilon_2 k_y ]^2 + (\upsilon_{+} k_y)^2 +(\upsilon_{1} k_x)^2 }
\label{energyIrr21}
\end{eqnarray}
\noindent
where the additional irradiation-induced band gap $\tilde{\Delta}_{\xi | s}(\lambda_0)$ takes the explicit form

\begin{eqnarray}
\tilde{\Delta}_{\xi | s}(\lambda_0) = \pm \sqrt{2c_0^2 \upsilon_1 \upsilon_{+} \Delta_{0}(1 - s r_{E} \xi) + c_0^4 \upsilon_1^2(\upsilon_2^2+\upsilon_{+}^2)}
\label{energyIrr22}
\end{eqnarray}

\noindent
and, $\lambda_0 = v_F e E_0 /(\hbar \omega^2) \ll 1 $, $\upsilon_{-}$,
$\upsilon_{+} $, $\upsilon_{1}$ and $\upsilon_{2}$  are as defined above as the Fermi velocities and the velocity correction terms. Furthermore, $\tau = \pm 1 $ labels the electron/hole states related to the conduction and the valence bands, while $ s = \pm 1 $ is the real spin index. The spin-orbit coupling gap is $\Delta_0 = 0.81 E^{(0)}$, $r_E = r = E_z/E_c$ is the relative value for the out-of-plane electric field, and $E_c$ stands for a critical field at which the band gap in 1T'-MoS$_2$ will be closed. It is important to notice that the energy dispersions are spin and valley polarized, i.e depend directly on indices $s$ and $\xi$. It is worth noting that the dressed states formed using linearly polarized dressing field are not studied in this paper. The model  Hamiltonian within the effective mass approximation for a single electron-hole pair in a  $\mathrm{1T'-MoS_{2}}$ double layer is given by

\begin{eqnarray}
\hat{H}_{0} = -
\frac{\hbar^{2}}{2m_{x}^{e}}\frac{\partial^{2}}{\partial x_{1}^{2}}
- \frac{\hbar^{2}}{2m_{y}^{e}}\left[\frac{\partial}{\partial y_{1}}
- p_{e}^{(0)}\right]^{2} -
\frac{\hbar^{2}}{2m_{x}^{h}}\frac{\partial^{2}}{\partial x_{2}^{2}}
- \frac{\hbar^{2}}{2m_{y}^{h}}\left[\frac{\partial}{\partial y_{2}}
- p_{h}^{(0)}\right]^{2} + V\left(\sqrt{r^{2}+D^{2}}\right)\ ,
\label{H02}
\end{eqnarray}
where $V\left(\sqrt{r^{2}+D^{2}}\right)$ is the potential energy for
electron-hole pair attraction, when the electron and hole are
located in two different 2D planes. In Eq.~(\ref{H02}), $m_{x}^{e}$, $m_{y}^{e}$, $m_{x}^{h}$, $m_{y}^{h}$, $p_{e}^{(0)}$, and $p_{h}^{(0)}$ are the constants, which can be obtained from Eq.~(\ref{energyIrr21}) using the long wavelength expansion.

\begin{equation}
E(k) = E_{0} + \frac{\hbar^2}{2 m^{*}_{x}} (k_x - k_{0,x})^2 + \frac{\hbar^2}{2 m^{*}_{y}} (k_y - k_{0,y})^2
\label{EK2}
\end{equation}
\noindent
Using the long wavelength expansion we are able to infer the following quantities

\begin{eqnarray}
m_{x}^{e/h} = \dfrac{ \sqrt{\Delta_0^2 (\xi - rs)^2 + \tilde{\Delta}^2_{\xi | s}(\lambda_0)}}{\upsilon_1^2}, \hspace{0.6cm} m_{y}^{e/h} = \frac{ [\Delta_0^2 (\xi - rs)^2 +\tilde{\Delta}^2_{\xi | s}(\lambda_0)]^{3/2}}{\Delta_0^2 \upsilon_{+}^2(\xi - rs)^2 +\tilde{\Delta}^2_{\xi | s}(\lambda_0)(\upsilon_2^2 + \upsilon_{+}^2)}
\label{effectiveMassFloquet}
\end{eqnarray}

\begin{eqnarray}
\tilde{p}_{e/h}^{(0)} =  \mp \frac{\left(\Delta_0^2 (\xi - rs)^2 + \tilde{\Delta}^2_{\xi | s}(\lambda_0)\right) \cdot \left({rs \upsilon_2 \Delta_0 - \upsilon \Delta_0 \xi + \upsilon_{-}\xi}\right) }{\upsilon_{+}^2 \Delta_0^2(\xi - rs)^2 + (\upsilon_2^2 + \upsilon_{+}^2)\tilde{\Delta}^2_{\xi | s}(\lambda_0) }
\end{eqnarray}

\noindent
The solutions to the center-of-mass and relative motion of the electron-hole pair are identical to those described in Section {I}.B, with the constants being the only difference.

\subsection{Binding energy}

\begin{figure}[H]
\centering
\includegraphics[width=0.6\textwidth]{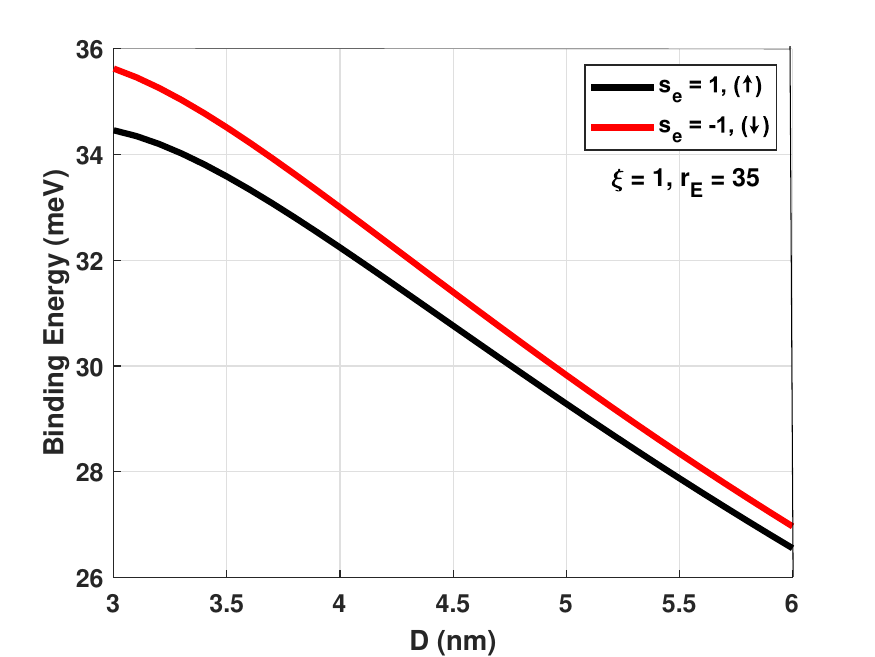}
\caption{(Color online)  The binding energies of excitons as functions of interlayer separation D for spin up (black) and spin down (red) electron-hole pairs. These results are for intravalley excitons at $\xi = 1$. The following combination of parameters is used for each of the binding energy calculations - $\mathcal{I}_{df} = 10^9$, $\omega = 10^{16}$ Hz, $r_E = 35$. We employed the binding energy definition of Eqn. (\ref{bindingEnergy}), along with the effective mass definitions of Eqn.(\ref{effectiveMassFloquet}), in the generation of these figures.}
\label{Floquet_BindingEnergy}
\end{figure}

In this section, we compute the binding energies of the excitons for chosen parameters. For our calculations, we confine our attention to $\xi = 1.0$. We perform calculations for both combinations of spins, spin up ($s_e = +1$) electrons and spin down holes,  and spin down electrons ($s_e = -1$) and spin up holes. In our calculations, it is worth noting that the mass of the holes is effectively that of an electron of opposite spin. In Fig.\ 6, the irradiation intensity $\mathcal{I}_{df} = 10^9$, the frequency of irradiation $\omega = 10^{16}$ Hz, and the relative value of the out-of-plane electric field $r_E = 35.0$. Our calculations show that spin down electrons-spin up hole pairs have a higher binding energy than spin up electrons-spin down hole pairs. These binding energies are also on a similar scale as the results outlined in Sec.\ \ref{sec1}.D in particular Figs. 3(b), 4(b) and 5(b).

\medskip
\par

 It is worth considering certain parameter restrictions and suggestions. This would enable us to obtain the binding energies for chosen interlayer eparation. Our choice of parameters leads to a higher binding energy for smaller interlayer separation. We find that choosing a $r_E$ of larger magnitude, leads to this effect. Additionally, the value of the gap must be real and positive.

\medskip
\par
\noindent
Furthermore, it is worth noting that changing $\omega$ and $\mathcal{I}_{df}$ does not induce an appreciable change in the effective masses. In accordance with prior results \cite{Floquet}, we find that the larger band gaps tend to be reduced by circularly polarized dressing field. The orientation and anisotropy for the constant-energy cut dispersions in 1T'-MoS$_2$ remains unchanged under circularly-polarized irradiation. The additional energy gap induced by the irradiation is contributed by the $\tilde{\Delta}$ term in~Eq.~(\ref{energyIrr22}).

\section{Collective excitations for spatially separated electrons and holes}
\label{sec4}

In this section, we examine the dilute limit for gases of electrons
and holes in a double layer of 1T$^\prime$-MoS$_2$. In this limit,
we consider dipolar {\bf Type A} and {\bf Type B} excitons. However,
the analysis holds true for all other combinations of types of
excitons, i.e., {\bf Type A} and {\bf Type C}, and {\bf Type B} and
{\bf Type C}. Henceforth, we will refer to {\bf Type A, B} and {\bf
C} excitons, simply as $A$, $B$ and $C$ excitons for brevity. In
this section we formulate
the sound velocity, which will determine whether the Landau
criterion for superfluidity is satisfied.

\medskip
\par
We expect that at $T = 0$ K almost all $A$ and $B$ excitons condense
into a BEC of $A$ and $B$ excitons. This allows us to assume the
formation of a binary mixture of BECs. {We will describe
this two-component weakly interacting Bose gas of excitons following
the procedure, described in
Refs.~\cite{SuperfluidDipolar,HighTempSuperOleg}.} We then examine
this mixture within the Bogoliubov approximation~\cite{Abrikosov}.
Only the interactions between the condensate and noncondensate
particles are considered, since we assume almost all the particles
belong to the BEC. The interactions between noncondensate particles
are neglected. This allows us to diagnolize the many particle
Hamiltonian.  This reduces the product of four operators in the
interaction term of the Hamiltonian by replacing it with a pair
consisting of a product of two operators \cite{LifPit}. The
condensate operators are replaced by numbers, and the resulting
Hamiltonian is quadratic with respect to the creation and
annihilation operators. Employing the Bogoliubov approximation
\cite{LifPit}, generalized for a two-component weakly interacting
Bose gas, we obtain the chemical potential~$\mu$ of the excitonic
system by minimizing $\hat{H}_0 - \mu \hat{N} $ with respect to the
2D concentration $n$~\cite{minimizeHmuN}, where $\hat{N}$ denotes
the number operator

\begin{equation}
\hat{N} = \sum_{\textbf{P}} a^{\dag}_{\textbf{p} A} a_{\textbf{p} A} + \sum_{\textbf{P}} a^{\dag}_{\textbf{p} B} a_{\textbf{p} B}
\label{numberOper}
\end{equation}
and $H_0$ is the Hamiltonian describing the particles in the condensate with zero momentum $\textbf{p} = 0$. In the Bogoliubov approximation, we assume $N = N_0$, $a^{\dag}_{\textbf{p}=0, A(B)} = \sqrt{N_{0A(B)}} e^{-i \Theta_{A(B)}}$, and $a_{\textbf{p}=0, A(B)} = \sqrt{N_{0A(B)}} e^{i \Theta_{A(B)}}$, where $N$ is the total number of all excitons in the condensate, $N_{0 A(B)}$ and $\Theta_{A(B)}$ are the number and phase for $A(B)$ excitons in the condensate. In the small momentum limit, $p = \hbar k$ when $\varepsilon_{(0)A}(P,\Theta) \ll G_{AA}$, and $\varepsilon_{(0)B}(P,\Theta) \ll G_{BB}$ we expand the spectrum of collective excitations $\varepsilon_{j}(p,\Theta)$ up to first order with respect to the momentum $p$ and obtain two sound modes of the collective excitations $\varepsilon_j(p,\Theta) = c_j(\Theta) p$, where $c_j (\Theta)$ is the sound velocity given by

\begin{equation}
c_j(\Theta) = \sqrt{\frac{G_{AA}}{2 M_{0A}(\Theta)}+\frac{G_{BB}}{2 M_{0B}(\Theta)}+(-1)^{j-1}\sqrt{\left(\frac{G_{AA}}{2 M_{0A}(\Theta)} -\frac{G_{BB}}{2 M_{0B}(\Theta)}\right)^2 + \frac{G^2_{AB}}{M_A(\Theta) M_B(\Theta)}}}
\label{soundV}
\end{equation}
\noindent
In the limit of large momenta when $\varepsilon_{(0)A}(P,\Theta) \gg G_{AA}$, and $\varepsilon_{(0)B}(P,\Theta) \gg G_{BB}$, we get two parabolic modes of collective excitations with the spectra $\varepsilon_1 (p,\Theta) = \varepsilon_{(0)A}(p,\Theta)$ and $\varepsilon_2 (p,\Theta) = \varepsilon_{(0)B}(p,\Theta)$. The Hamiltonian $\hat{H}_{\text{col}}$ of collective excitations, corresponding to two branches of the spectrum, in the Bogoliubov approximation for the entire two-component anisotropic system is given by

\begin{equation}
\hat{H}_{\text{col}} =\sum_{\textbf{P} \neq 0} \varepsilon_1 (P,\Theta) \alpha^{\dag}_{1\textbf{p}} \alpha_{1\textbf{p}} + \sum_{\textbf{P}\neq 0} \varepsilon_2 (P,\Theta) \alpha^{\dag}_{2\textbf{p}} \alpha_{2\textbf{p}}  \  ,
\label{Hcoll}
\end{equation}
\noindent where $\alpha^{\dag}_{j\textbf{P}}$ and
$\alpha_{j\textbf{P}}$ are the creation and annihilation Bose
operators for the quasiparticles with the energy dispersion
corresponding to the $j$th mode of the spectrum of the collective
excitations. For simplicity, we consider the specific case when the
densities of $A$ and $B$ excitons are the same as $n_A = n_B = n/2$.
Following the standard procedure for calculations outlined in the Supplemental Materials, Section II in~\cite{SuppMat} we obtain the spectrum of collective excitations

\begin{equation}
\varepsilon_j(P,\Theta) = \sqrt{\frac{\omega^2_A(P,\Theta)+\omega^2_B(P,\Theta) +(-1)^{j-1}\sqrt{\left[\omega^2_A(P,\Theta) -\omega^2_B(P,\Theta)\right]^2 + 4 g^2 n^2\varepsilon_{(0)A}(P,\Theta)\varepsilon_{(0)B}(P,\Theta)}}{2}}
\label{spectrumJ2}
\end{equation}
\noindent
and the sound velocity at $n_A = n_B = n/2$ is obtained as
\noindent
\begin{equation}
c_j(\Theta) = \sqrt{\frac{gn}{2}\left(\frac{1}{2M_{0A}(\Theta)} + \frac{1}{2M_{0B}(\Theta)} + (-1)^{j-1} \sqrt{\left(\frac{1}{2M_{0A}(\Theta)} - \frac{1}{2M_{0B}(\Theta)}\right)^2+ \frac{1}{M_{0A}(\Theta)M_{0B}(\Theta)}}\right)}
\label{soundVelEq}
\end{equation}
\noindent
It follows from Eq.~(\ref{soundVelEq}) that there is only one nonzero sound velocity at $n_A = n_B = n/2$ given by

\begin{equation}
c_S(\Theta) = \sqrt{\frac{gn}{2}\left(\frac{1}{M_{0A}(\Theta)} + \frac{1}{M_{0B}(\Theta)}\right)}
\label{soundVelEq2}
\end{equation}

It is also worth noting that the large interlayer separation $D$, between the double layer allows us to neglect the exchange interactions in the electron-hole system. This is caused by a low tunneling probability , caused by the shielding of the dipole-dipole interaction by the dielectric (h-BN) that separates the layers in the double layer \cite{HighTempSuperOleg}.

\section{ Superfluidity}
\label{sec5}

In this section, we determine the criterion for which a weakly
interacting Bose gas of dipolar excitons can form a superfluid.
Since at low momenta the energy spectrum of quasiparticles of a
weakly interacting gas of excitons is soundlike, this system
satisfies the Landau criterion for superfluidity \cite{LifPit}. The
critical velocity for superfluidity is given by $v_c =
\text{min}(c_1(\Theta),c_2(\Theta))$ since the quasiparticles are
created at velocities exceeding that of sound for the lowest mode of
the quasiparticle dispersion. Additionally, it is angular dependent.
The density of the superfluid   $\rho_s(T)$ is defined as $\rho_s(T)
= \rho - \rho_n(T)$, where $\rho = M_A n_A + M_B n_B$, is the total
2D density of the system and $\rho_n(T)$ is the density of the
normal component. We define the normal component $\rho_n(T)$ in the
usual way \cite{nanomatGabriel}. Assume that the excitonic system
moves with velocity \textbf{u}, which means that the superfluid
component has velocity \textbf{u}. At finite temperatures $T$,
dissipating quasiparticles will emerge in this system. Since their
density is small at low temperatures, one may assume that the gas of
quasiparticles is an ideal Bose gas. {We will obtain the
density of the superfluid component in the anisotropic system
following the procedure, described in Ref. \cite{phosphoreneOGR}. To
calculate the superfluid component density, we begin by defining the
mass current \textbf{J} for a Bose gas for quasiparticles in the
frame of reference where the superfluid component is at rest by

\begin{equation}
\textbf{J} = s \int \frac{d^2P}{(2 \pi \hbar)^2} \textbf{P}\left[f[\varepsilon_1(P,\Theta) -\textbf{P}\textbf{u}]+f[\varepsilon_2(P,\Theta)-\textbf{P}\cdot\textbf{u}]\right]
\label{massCurrent}
\end{equation}
\noindent
where $f[\varepsilon_1(P,\Theta)] = \{ \text{exp}[\varepsilon_1(P,\Theta)/(k_B T)] -1\}^{-1} $ and $f[\varepsilon_2(P,\Theta)] = \{ \text{exp}[\varepsilon_2(P,\Theta)/(k_B T)] -1\}^{-1}$ are the Bose-Einstein distribution functions for the quasiparticles with the angle dependent dispersions $\varepsilon_1(P,\Theta)$ and $\varepsilon_2(P,\Theta)$ respectively, $s$ is the spin degeneracy factor, and $k_B$ is the Boltzmann constant. Expanding the expression under the integral in terms of $\textbf{P}\textbf{u}/k_B T$ and restricting ourselves to the first-order term, we obtain

\begin{equation}
\textbf{J} = -\frac{s}{k_B T} \int \frac{d^2P}{(2 \pi \hbar)^2} \textbf{P} (\textbf{P}\cdot \textbf{u}) \left(\frac{\partial f [\varepsilon_1(P,\Theta)]}{\partial \varepsilon_1(P,\Theta)} +\frac{\partial f [\varepsilon_2(P,\Theta)]}{\partial \varepsilon_2(P,\Theta)}\right)
\label{MassCurrentFirstOrder}
\end{equation}
\noindent
The normal density $\rho_n$ in the anisotropic system takes tensor form. We define the tensor elements for the normal component density $\rho^{(ij)}_n(T)$ by

\begin{equation}
J_i = \rho^{(ij)}_n(T) u_j
\label{MassCurrentTensor}
\end{equation}
\noindent
where $i,j$ denote either $x,y$
component.  Assuming that the vector \textbf{u} is parallel to the $OX$ axis and has the same direction as this axis, we have $\textbf{u} = u_x \textbf{i}$ and $\textbf{P} = P_x \textbf{i} + P_y \textbf{j}$ and \(
\textbf{P} \cdot \textbf{u} = P_x u_x\), \(
\textbf{P} (\textbf{P} \cdot \textbf{u}) = P_x^2 u_x \textbf{i} + P_x P_y u_x \textbf{j}\),
where \textbf{i} and \textbf{j} are unit vectors in the $x$ and $y$ directions, respectively. We then obtain

\begin{equation}
J_x = -\frac{s}{k_B T} \int_{0}^{\infty} dP \frac{P^3}{(2 \pi \hbar)^2} \int_{0}^{2 \pi} d\Theta \left(\frac{\partial f [\varepsilon_1(P,\Theta)]}{\partial \varepsilon_1(P,\Theta)} +\frac{\partial f [\varepsilon_2(P,\Theta)]}{\partial \varepsilon_2(P,\Theta)}\right) \cos^2 \Theta \ u_x
\label{masscurrentX}
\end{equation}
\noindent
Using the definition of the density for the normal component from Eq.~(\ref{MassCurrentTensor}), we obtain

\begin{equation}
\rho^{(xx)}_n(T) = \frac{s}{k_B T} \int_{0}^{\infty} dP \frac{P^3}{(2 \pi \hbar)^2} \int_{0}^{2 \pi} d\Theta \left( \frac{\text{exp}[\varepsilon_1(P,\Theta)/(k_BT)]}{\{ [\text{exp}[\varepsilon_1(P,\Theta)/(k_BT)] -1\}^2} + \frac{\text{exp}[\varepsilon_2(P,\Theta)/(k_BT)]}{\{ [\text{exp}[\varepsilon_2(P,\Theta)/(k_BT)] -1\}^2} \right) \cos ^2 \Theta
\label{rhoxx}
\end{equation}
\noindent
Furthermore, from Eq.~(\ref{MassCurrentFirstOrder}), we can also obtain

\begin{align}
J_y &= -\frac{s}{k_B T} \int_{0}^{\infty} \frac{d^2P}{(2 \pi \hbar)^2} P_x P_y \left(\frac{\partial f [\varepsilon_1(P,\Theta)]}{\partial \varepsilon_1(P,\Theta)} +\frac{\partial f [\varepsilon_2(P,\Theta)]}{\partial \varepsilon_2(P,\Theta)}\right) u_x \\
&= \frac{s}{k_BT} \int_{0}^{\infty} dP \frac{P^3}{(2\pi \hbar)^2} \int_{0}^{2 \pi} d \Theta \left( \frac{\text{exp}[\varepsilon_1(P,\Theta)/(k_BT)]}{\{ [\text{exp}[\varepsilon_1(P,\Theta)/(k_BT)] -1\}^2} + \frac{\text{exp}[\varepsilon_2(P,\Theta)/(k_BT)]}{\{ [\text{exp}[\varepsilon_2(P,\Theta)/(k_BT)] -1\}^2} \right) \cos \Theta \sin \Theta \ u_x
\label{masscurrentY}
\end{align}

The integral in Eq.~(\ref{masscurrentY}) is zero, since the $\Theta$ integral over the period of the integrand vanishes.  Therefore, one determines that $\rho^{(xy)}_n = 0$. Now, assuming that the vector \textbf{u} is parallel to the $OY$ axis, we obtain analogously the following relations:

\begin{align}
\rho^{(yy)}_n (T) &= \frac{s}{k_B T} \int_{0}^{\infty} dP \frac{P^3}{(2 \pi \hbar)^2} \int_{0}^{2 \pi} d\Theta \left( \frac{\text{exp}[\varepsilon_1(P,\Theta)/(k_BT)]}{\{ [\text{exp}[\varepsilon_1(P,\Theta)/(k_BT)] -1\}^2} + \frac{\text{exp}[\varepsilon_2(P,\Theta)/(k_BT)]}{\{ [\text{exp}[\varepsilon_2(P,\Theta)/(k_BT)] -1\}^2} \right) \sin^2 \Theta \\
\rho^{(yx)}(T) &= 0
\label{OYEqns}
\end{align}
\noindent
By defining the tensor of the concentration of the normal component as the linear response of the flow of quasiparticles on the external velocity as $n^{(ij)}_n =\rho^{(ij)}_n/M_i$, one obtains

\begin{align}
n^{(xx)}_n(T) &= \frac{s}{k_B M_x T} \int_{0}^{\infty} dP \frac{P^3}{(2 \pi \hbar)^2} \int_{0}^{2 \pi} d\Theta \left( \frac{\text{exp}[\varepsilon_1(P,\Theta)/(k_BT)]}{\{ [\text{exp}[\varepsilon_1(P,\Theta)/(k_BT)] -1\}^2} + \frac{\text{exp}[\varepsilon_2(P,\Theta)/(k_BT)]}{\{ [\text{exp}[\varepsilon_2(P,\Theta)/(k_BT)] -1\}^2} \right) \cos ^2 \Theta \\
n^{(yy)}_n(T) &= \frac{s}{k_B M_y T} \int_{0}^{\infty} dP \frac{P^3}{(2 \pi \hbar)^2} \int_{0}^{2 \pi} d\Theta \left( \frac{\text{exp}[\varepsilon_1(P,\Theta)/(k_BT)]}{\{ [\text{exp}[\varepsilon_1(P,\Theta)/(k_BT)] -1\}^2} + \frac{\text{exp}[\varepsilon_2(P,\Theta)/(k_BT)]}{\{ [\text{exp}[\varepsilon_2(P,\Theta)/(k_BT)] -1\}^2} \right) \sin^2 \Theta
\end{align}

\begin{equation}
n^{(xy)}(T) = 0 , \hspace{0.2cm} n^{(yx)}(T) = 0
\end{equation}
\noindent
The linear response of the flow of quasiparticles $\textbf{J}_{qp}$ with respect to the external velocity at any angle measured from the $OX$ direction is given in terms of the angle-dependent concentration for the normal component $\tilde{n}_n (\Theta, T)$ as

\begin{eqnarray}
|\textbf{J}_{{qp}}| &=& |n^{(xx)}_n (T) \ u_x \ \textbf{i} + n^{(yy)}_n (T) \  u_y \ \textbf{j}|
\nonumber\\
&=&\sqrt{[n^{(xx)}_n(T)]^2u^2 \cos^2 \Theta + [n^{(yy)}_n(T)]^2u^2 \sin^2 \Theta}
= \tilde{n} (\Theta, T) u  \  ,
\label{Jqp}
\end{eqnarray}
where the concentration of the normal component  is

\begin{equation}
\tilde{n}_n (\Theta, T) =  \sqrt{[n^{xx}_n(T)]^2  \cos^2 \Theta + [n^{yy}_n(T)]^2 \sin^2 \Theta}  \  .
\label{normalComp}
\end{equation}
\noindent
From Eq.~(\ref{normalComp}) it follows that $n^{(xx)}_n = \tilde{n}_n (\Theta = 0)$ and $n^{(yy)}_n = \tilde{n}_n (\Theta = \pi/2)$. We can then rewrite Eq.~(\ref{normalComp}) in the following form:

\begin{equation}
\tilde{n}_n (\Theta, T) =  \sqrt{\frac{[n^{xx}_n(T)]^2+[n^{yy}_n(T)]^2}{2} + \frac{([n^{xx}_n(T)]^2-[n^{yy}_n(T)]^2) \cos(2\Theta)}{2}}
\label{normalCompRewrite}
\end{equation}
\noindent
We define the angle-dependent concentration of the superfluid component $\tilde{n}_s(\Theta, T)$ by

\begin{equation}
\tilde{n}_s(\Theta, T) = n - \tilde{n}_n(\Theta, T)   \  ,
\label{superfluidNormalN}
\end{equation}
\noindent
where $n$ is the total concentration of the excitons. The mean-field critical temperature $T_c(\Theta)$ of the phase transition related to the occurrence of superfluidity in the direction with the angle $\Theta$ relative to the $x$ direction is determined by the condition
\( \tilde{n}_n(\Theta, T) = n  \).

\subsection{Superfluidity for the soundlike spectrum of collective excitations}

In a straightforward way, it follows from  Eq.~(\ref{soundV}) that for $j = 2$ the sound velocity vanishes. Therefore, we only take into account the spectrum of collective  excitations for $j = 1$. According to Ref.\ \cite{LandauStat}, it is clear that we need a finite sound velocity for superfluidity. Since the branch of collective excitations at zero sound velocity corresponds to zero energy of the quasiparticles (which means that no quasiparticles are created at zero sound velocity), this branch does not lead to dissipation of energy, thereby resulting in finite velocity and does not affect the Landau critical velocity. The weakly interacting gas of dipolar excitons, satisifies the Landau criterion for superfluidity since for small momenta, the energy spectrum of the quasiparticles in the weakly interacting gas of dipolar excitons at $j = 1$ is soundlike with sound velocity

\begin{equation}
c_1 (\Theta) = \sqrt{\frac{g n_A}{M_{OA}(\Theta)} + \frac{g n_B}{M_{OB}(\Theta)}} \  .
\end{equation}
\noindent
Clearly the ideal Bose gas has no branch with non-zero sound velocity, thus not demonstrating superfluidity. At low temperatures, the two-component system of dipolar excitons exhibits superfluidity due to exciton-exciton interactions. With this new knowledge, we can again return to our definition for the mass current in Eq.~(\ref{MassCurrentFirstOrder}) rewritten as

\begin{equation}
\textbf{J} = {s} \int \frac{d^2P}{(2 \pi \hbar)^2} \textbf{P} f [\varepsilon_1 (P,\Theta) - \textbf{P}\cdot \textbf{u}]  \  ,
\label{MassCurrentFirstOrderRevisit}
\end{equation}
\noindent
where $s = 16 $ is the spin degeneracy factor, $f [\varepsilon_1 (p)]$ is the Bose-Einstein distribution function for the quasiparticles with dispersion $\varepsilon_1 (p)$ and $k_B$ is the Boltzmann constant. Expanding this to first order with respect to $ \textbf{P} \cdot \textbf{u} / (k_B T)$, we obtain

\begin{equation}
\textbf{J} = -s \frac{\textbf{u}}{2} \int \frac{d^2P}{(2 \pi \hbar)^2} P^2 \frac{\partial f [\varepsilon_1(P,\Theta)]}{\partial \varepsilon_1(P,\Theta)} \ .
\label{MassCurrentFirstOrderRevisitExpanded}
\end{equation}
\noindent
The density of the normal component $\rho_n$, in the moving weakly-interacting Bose gas of dipolar excitons is defined as
\(
\textbf{J} = \rho_n \textbf{u}
\  . \)
Therefore,
\begin{equation}
\rho_n = -\frac{s}{2} \int \frac{d^2P}{(2 \pi \hbar)^2} P^2 \frac{\partial f [\varepsilon_1(P,\Theta)]}{\partial \varepsilon_1(P,\Theta)}
\label{rhoNrevisited}
\end{equation}
\noindent
By substituting the sound velocity $\varepsilon_1 (P,\Theta) = c_1(\Theta) P$ into Eq.~(\ref{rhoNrevisited}), we obtain

\begin{equation}
\rho_n (T) = \frac{3s \zeta(3)}{2 \pi \hbar^2 c_1^4 (\Theta)} k_B^3 T^3
\label{rhoNT}
\end{equation}
\noindent
The mean field critical temperature $T_c$ of the phase transition at which the superfluidity occurs, implying neglecting the interaction between the quasiparticles, is obtain from the condition $\rho_s (T_c) = 0$:

\begin{equation}
\rho_n (T_c,\Theta) = \rho = M_{OA} n_A + M_{OB} n_B
\label{rhoCritTemp}
\end{equation}
\noindent
At low temperatures $k_B T \ll M_{A(B)} c_1(\Theta)^2$, by substituting Eqn.~(\ref{rhoCritTemp}) into Eqn.~(\ref{rhoNT}) one derives

\begin{equation}
T_c (\Theta) = \left[\frac{2 \pi \hbar^2 \rho c_1^4 (\Theta)}{3 \zeta(3) s k_B^3}\right]^{1/3} \ .
\label{rhoCritTemp2}
\end{equation}
\noindent
In this paper, we have obtained the mean-field critical temperature $T_c$ of the phase transition at which superfluidity appears without claiming BEC in a 2D system at finite temperature.

\section{Results and Discussion}
\label{sec6}

\subsection{Computation of sound velocity as function of different parameters}

We now demonstrate  numerically the dependence of the sound velocity and the critical temperature on the angle $\Theta$ and the chose parameters for an experiment such as the interlayer separation $D$ and the relative value of the out-of-plane electric field $\alpha$. It is worth noting that the formalism outlined above can apply to any pairs of excitons - A and B, A and C, and B and C exciton pairs. For the masses in this section, we will be looking at the non-Floquet engineered masses specified via Eqn.\ (\ref{effectiveMassesXY}).

\medskip
\par
In Fig.\ (\ref{SoundVelocityVariance1}),  the sound velocity $c_1$ is plotted as a function of the angle~$\Theta$. The sound velocity is seen to have a maximum at $\pi/2$ and $3 \pi/2$ and is minimum at $\pi$ and $2 \pi$. The sound velocity oscillates for all pairs of excitons and has maxima at around $1.5 \times 10^5$ m/s for all pairs of excitons for the parameters we have chosen below.  The minima are about $8.0 \times 10^4$ m/s.

\begin{figure}[H]
\centering
\includegraphics[width=0.52\textwidth]{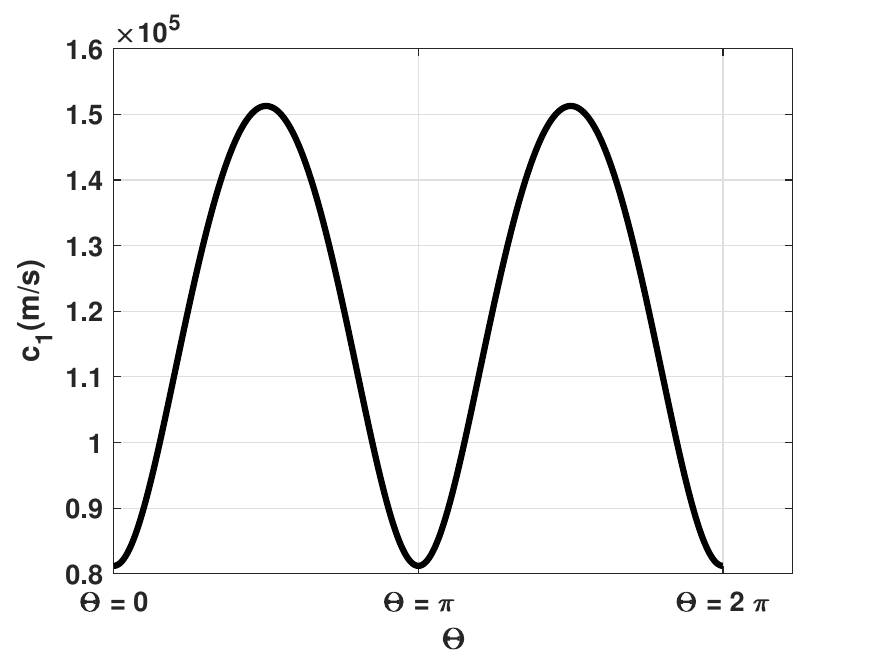}
\caption{The dependence of the sound velocity $c_1$ as a function of $\Theta$ for the pairing of Type A and B excitons. The difference in the sound velocity for different pairings of excitons, as a function of $\Theta$ is negligibly small. The parameters that have been kept constant for this are $n = 1$~x~$10^{11} $cm$^{-2}$, $D = 25$ nm and $\alpha = 35$. We employed the definition of the sound velocity in Eqn. (\ref{soundVelEq2}) in the generation of these figures.}
\label{SoundVelocityVariance1}
\end{figure}
\noindent
In Fig.~(\ref{SoundVelocityVariance2}) below, we plot the dependence of the sound velocity $c_1$ as a function of the interlayer separation $D$ (nm). It is worth noting that higher values of interlayer separation lead to higher sound velocities for all pairs of excitons; It is worth noting that at our recommended number h-BN layers - 10 layers at 3.3 $\text{\r{A}}$ per layer for an interlayer separation of 3.3 nm - the sound velocity is around 2 x 10$^4$ m/s.

\begin{figure}[H]
\centering
\includegraphics[width=0.52\textwidth]{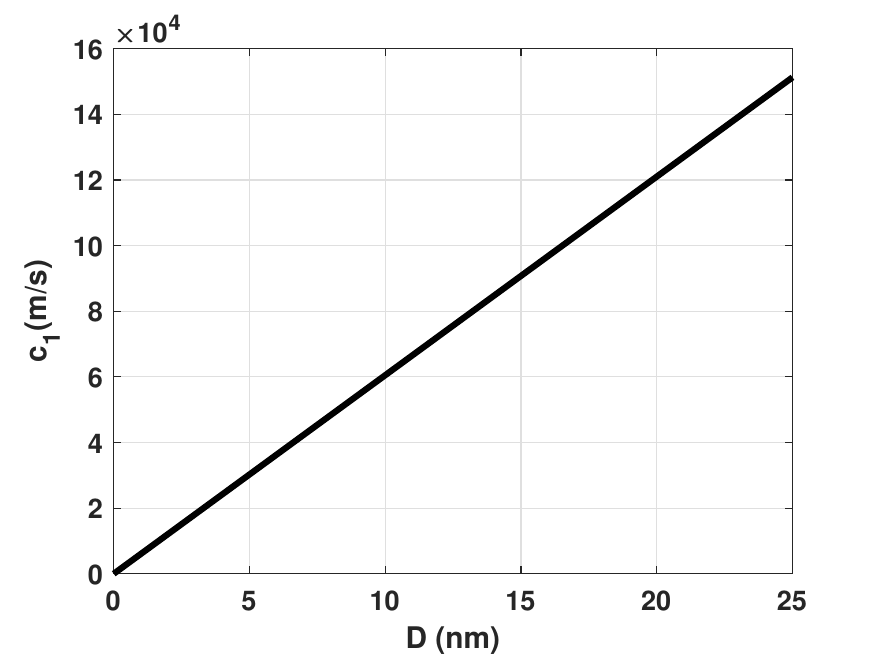}
\caption{The dependence of the sound velocity $c_1$ as a function of $D$ for different pairings of excitons. The difference in the sound velocity for different pairings of excitons, as a function of $D$ is negligibly small. The parameters that have been kept constant for this are $n = 1$~x~$10^{11} $cm$^{-2}$, $\alpha= 35$ and $\Theta = \pi/2$. We employed the definition of the sound velocity in Eqn. (\ref{soundVelEq2}) in the generation of these figures.}
\label{SoundVelocityVariance2}
\end{figure}

\medskip
\par
In Fig.\ \ref{SoundVelocityVariance3}, we plot the sound velocity as a function of the relative value of the out-of-plane electric field $\alpha$. Here we notice some distinctly different behavior for the exciton-exciton pairs, specifically in the range $0 < \alpha < 1$. For A and B, and A and C exciton-exciton pairs, the sound velocity initially has a base value which increases from $0$ to $1.0$. When $\alpha= 1.0$, there is no calculated sound velocity because at this critical value of $\alpha$, the A excitons have zero center-of-mass, as demonstrated in Fig. 1(b). The sound velocity is monotonically
decreasing beyond $\alpha > 1$. For the pairing of B and C excitons, the sound velocity decreases from $0<\alpha<1$, before steadily declining at larger values of~$\alpha$. In Fig. 9 below this behaviour is not captured appreciably, and at about $\alpha > 2$ the sound velocity plateaus off to a fairly constant value.

\begin{figure}[H]
\centering
\includegraphics[width=0.52\textwidth]{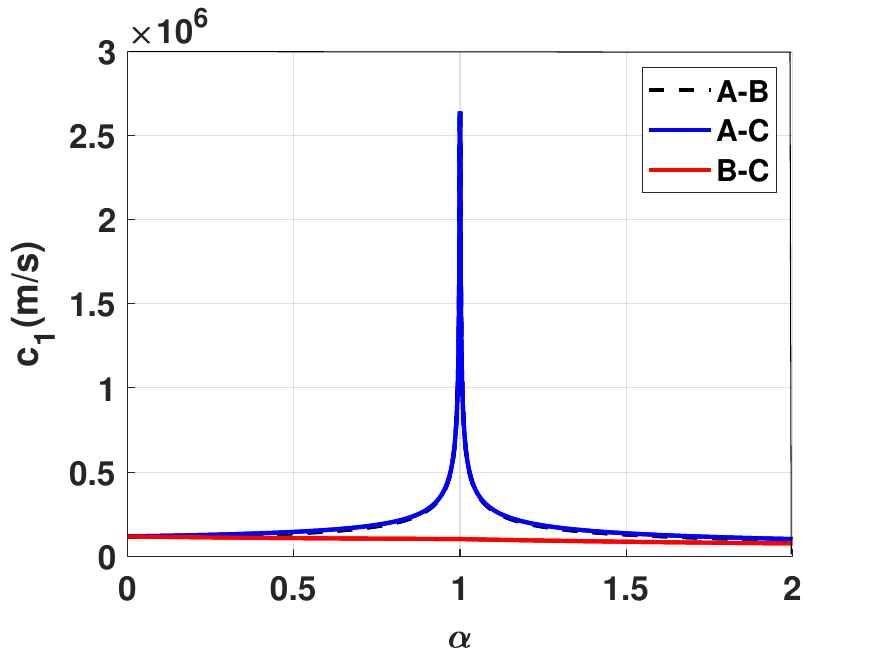}
\caption{(Color online)  The dependence of the sound velocity $c_1$ as a function of $\alpha$ for different pairings of excitons. The parameters that have been kept constant for this are $n = 1$~x~$10^{11} $cm$^{-2}$, $D = 3.3$ nm and $\Theta = \pi/2$. We employed the definition of the sound velocity in Eqn. (\ref{soundVelEq2}) in the generation of these figures.}
\label{SoundVelocityVariance3}
\end{figure}

\subsection{The critical temperature}

We now turn our attention to calculating the critical temperature as functions of various chosen parameters. We begin by investigating  the dependence of the critical temperature as a function of the angle $\Theta$ shown in Fig.\ (\ref{CritTempVariance1}). As expected, the behavior here is also oscillatory, displaying similar maxima and minima as the dependence of the sound velocity on the angle $\Theta$. The different pairs of excitons demonstrate negligibly different critical temperatures. When zoomed in the A and C exciton-exciton pairs display generally higher critical temperatures, and B and C exciton-exciton pairs display generally lower critical temperatures, but notably the difference is a very small percentage of their mean value.

\begin{figure}[H]
\centering
\includegraphics[width=0.52\textwidth]
{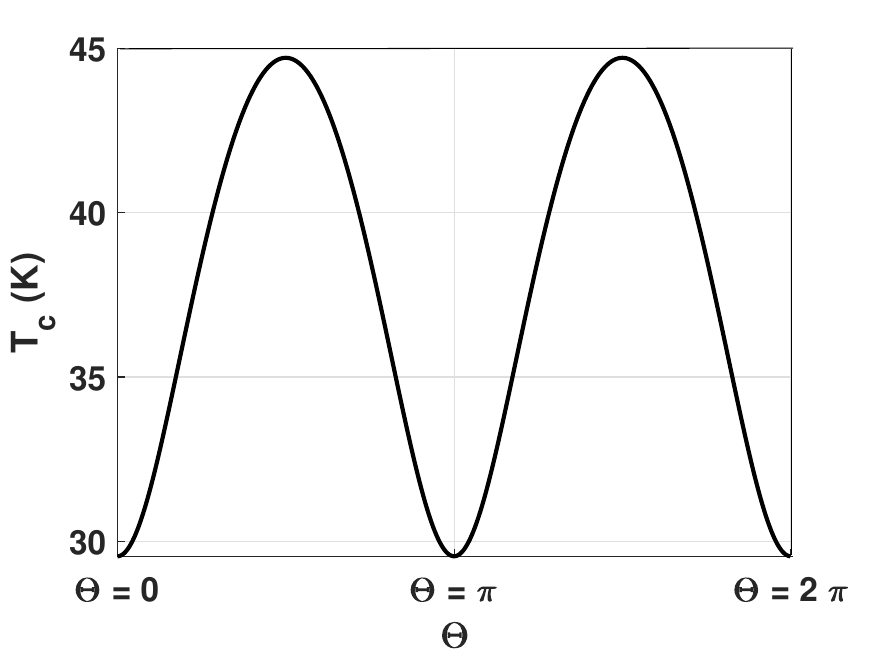}
\caption{The dependence of the critical temperature as a function of $\Theta$ for different pairing of A and B excitons. The difference in the critical temperature for different pairings of excitons, as a function of $\Theta$ is negligibly small. The parameters that have been kept constant for this are $n = 20$ x $10^{11} $cm$^{-2}$, $D = 3.3$ nm and $\alpha = 35$. We employed the definition of the critical temperature, $T_c$ in Eqn. (\ref{rhoCritTemp2}) in the generation of these figures.}
\label{CritTempVariance1}
\end{figure}

We plot the critical temperature as a function of the interlayer separation $D$. We chose the range $3.0\ \textnormal{nm}~\leq~D~\leq~6.0\ \textnormal{nm}$ since this is a range of interest for a feasible number of h-BN layers in an experiment. This smaller range also helps us see with greater resolution the contrast between the different pairings of exciton-exciton pairings. Larger interlayer separations lead to higher critical temperatures.

\begin{figure}[H]
\centering
\includegraphics[width=0.52\textwidth]
{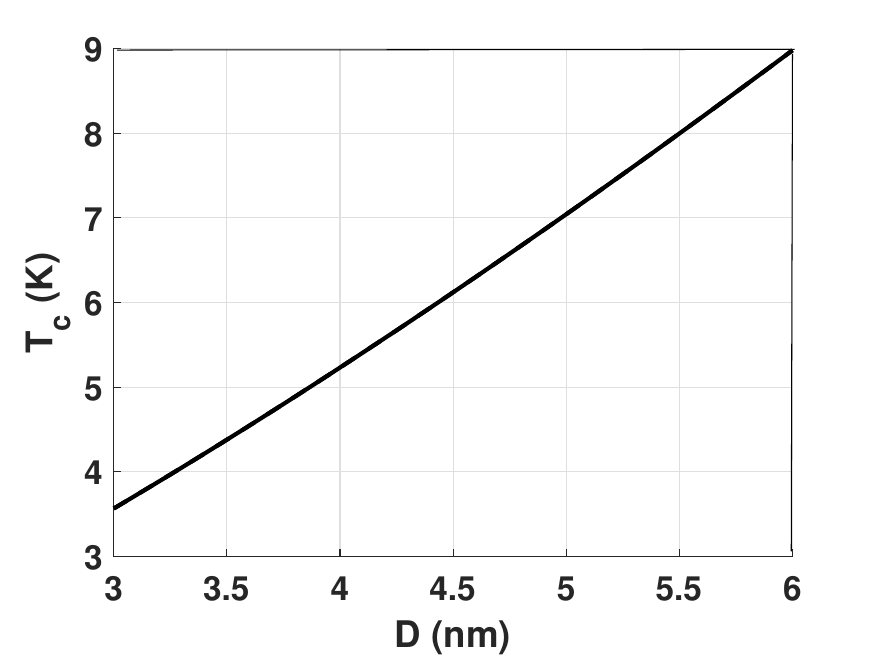}
\caption{The dependence of the critical temperature as a function of $D$ for different pairings of excitons. The difference in the critical temperature for different pairings of excitons, as a function of $D$ is negligibly small. The parameters that have been kept constant for this are $n = 1$ x $10^{11} $cm$^{-2}$, $\Theta = \pi/2$ and $\alpha = 35$. We employed the definition of the critical temperature, $T_c$ in Eqn. (\ref{rhoCritTemp2}) in the generation of these figures.}
\label{CritTempVariance2}
\end{figure}

We then investigate the critical temperature as a function of $\alpha$. Our results are presented in Fig.\  \ref{CritTempVariance3}. The critical temperature is increased in the range $0< \alpha <1.0$ for A-B and A-C exciton-exciton pairs. At $\alpha = 1.0$, the critical temperature is notably unable to be calculated because of the zero center-of-mass for A excitons. For all pairs of excitons, the critical temperature is decreased beyond $\alpha > 1.0$. Since using a large values of $\alpha$ is favorable for a larger binding energy, and thus stability and lifetime of the exciton,  we need to balance the choice of the parameter with the diminishing of the critical temperature $T_c$.

\begin{figure}[H]
\centering
\includegraphics[width=0.52\textwidth]{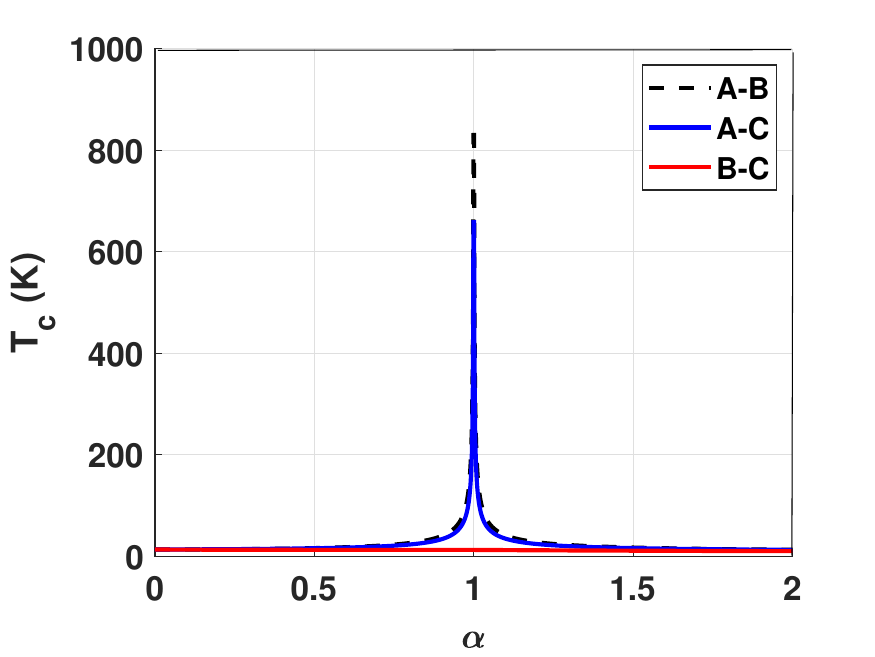}
\caption{(Color online)  The dependence of the critical temperature as a function of $\alpha$ for different pairings of excitons. The parameters that have been kept constant for this are $n = 1$ x $10^{11} $cm$^{-2}$, $\Theta = \pi/2$ and $D = 3.3$ nm. We employed the definition of the critical temperature, $T_c$ in Eqn. (\ref{rhoCritTemp2}) in the generation of these figures.}
\label{CritTempVariance3}
\end{figure}

It is worth calculating the critical temperature for the set of experimental parameters we recommend. These are as a function of the exciton concentration $n$, where $n = n_A/2 = n_B/2$. For this calculation, we set the parameters as follows - $\alpha = 35.0$, $D = 3.3$ nm (corresponding to ten layers of h-BN) and $\Theta = \pi/2$. Our results are presented in Fig.\ \ref{CritTempVariance4}. All combinations of exciton-exciton pairs lead to fairly similar critical temperatures for different concentrations of excitons. Zooming in, one can see that the A-C exciton-exciton pair has a higher critical temperature than the other two combinations, while B-C has a lower critical temperature, for a chosen set of parameters. Since we consider the dilute limit, we demand that the average distance between the excitons is much larger than the interlayer separation. Mathematically, this corresponds to $n \ll 1/(\pi D^2)$.

\begin{figure}[H]
\centering
\includegraphics[width=0.52\textwidth]{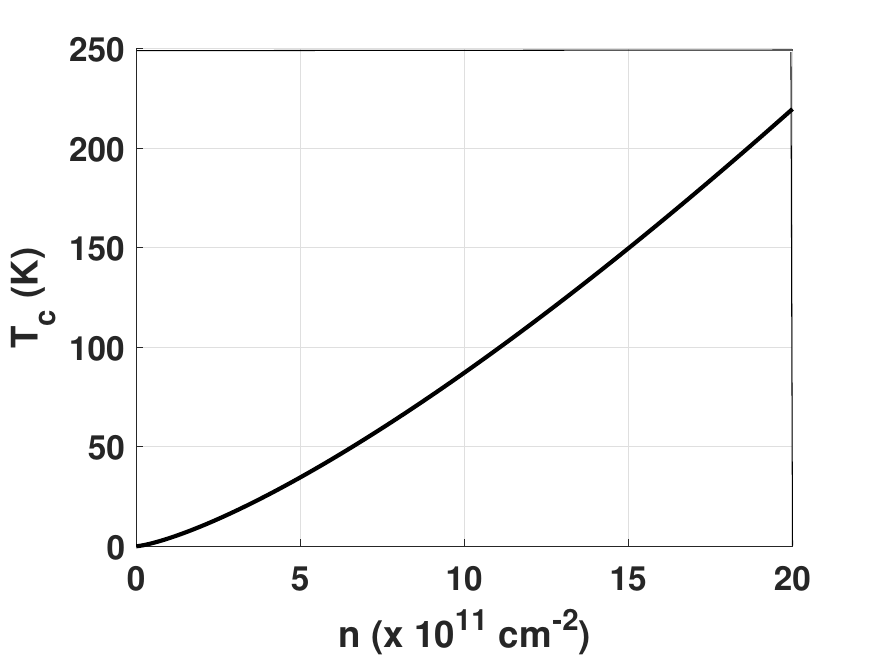}
\caption{The dependence of the critical temperature as a function of $n$ for different pairings of excitons. The difference in the critical temperature for different pairings of excitons, as a function of $n$ is negligibly small. The parameters that have been kept constant for this are $\alpha = 35$, $\Theta = \pi/2$ and $D = 3.3$ nm. We employed the definition of the critical temperature, $T_c$ in Eqn.~(\ref{rhoCritTemp2}) in the generation of these figures.}
\label{CritTempVariance4}
\end{figure}

It is not necessary to restrict our attention to the condition $n = n_A/2 = n_B/2$, for any two pairs of excitons. As such, in Fig.\ 14, we investigate the dependence of $T_c$ as a     function of the exciton densities. We notice that an increase in exciton density for both types of excitons leads to a higher critical temperature. We note that the increase in concentration to a value of $n_A = n_B = 2.0 \times 10^{12}$ cm$^{-2}$, leads to a critical temperature of around $T_c = 250$ K, which holds true for any combination of any two excitons of A, B or C. It is worth noting that in the pairing of B and C excitons, the concentration of B excitons has a more pronounced effect on the $T_c$.

\begin{figure}[h!]

\centering \subfigure(a){
\includegraphics[width=0.46\textwidth]{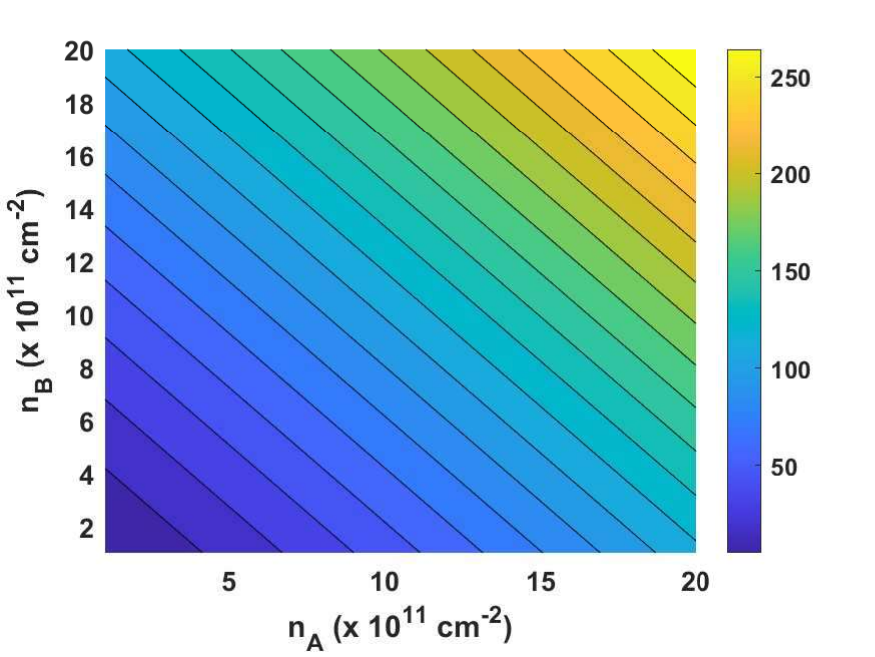}
} \subfigure(b){
\includegraphics[width=0.46\textwidth]{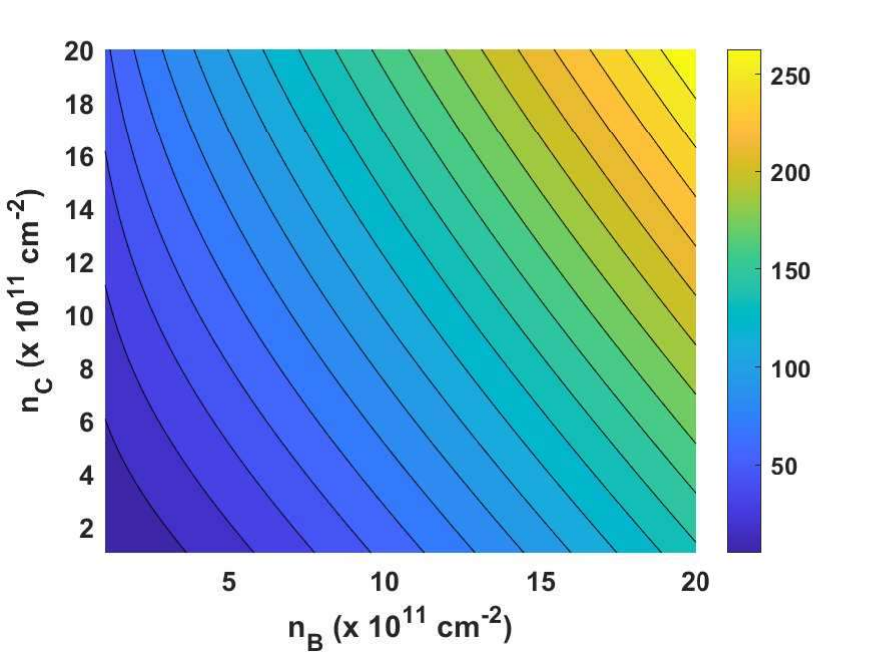}
} \subfigure(c){
\includegraphics[width=0.46\textwidth]{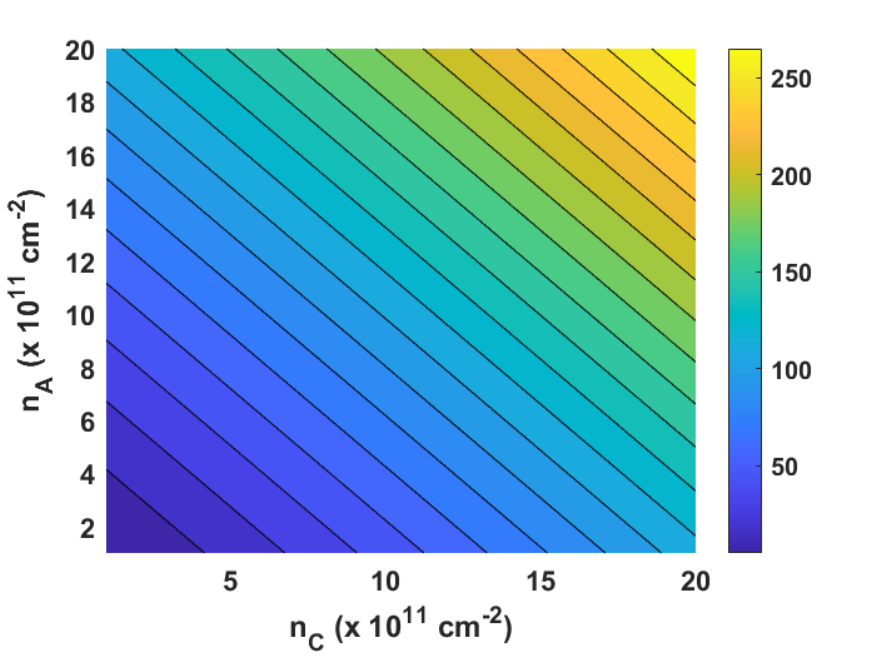}
}
\caption{(Color online)  The dependence of the critical temperature as a function of the exicton densities for different pairings of excitons. The parameters that have been kept constant for this are $\alpha = 35$, $\Theta = \pi/2$ and $D = 3.3$ nm. The colorbar on the side is the critical temperature $T_c$ in Kelvins. We employed the definition of the critical temperature, $T_c$ in Eqn. (\ref{rhoCritTemp2}) in the generation of these figures.}
\label{Fig1}
\end{figure}

\newpage

\section{Conclusions and discussion}
\label{sec7}

In this paper, we have investigated {the binding
energies, wave functions, collective properties and} superfluidity
of dipolar excitons {in a double layer of massive
anisotropic tilted Dirac systems} irradiated by
circularly polarized light. {For our calculations as an
example we have considered  a double layer of  1T$^\prime$-MoS$_2$.}
The binding energy of the three different types of excitons has been
obtained by employing the harmonic oscillator approximation. It is
found that using a large value of a perpendicular electric field
$\alpha$ leads to a large binding energy at a small enough
interlayer separation $D$ to use an experimentally reasonable number
of layers of h-bN dielectric. It is also found that using Floquet
engineering of the energy bands leads to no appreciable change in
the effective mass of the electrons or holes. However, it is worth
noting that it is unclear if using a larger $\lambda_0$ value,
corresponding to a larger irradiation intensity to frequency ratio
would lead to an appreciable increase in the effective mass. To study
this, a second-order or higher expansion of the energy dispersion
would be necessitated. However, as seen in Fig.\ 6, the binding
energy of the excitons is very close in magnitude to the binding
energy without any circularly polarized dressing field applied.

\medskip
\par
We note that for dipolar excitons in 1T$^\prime$-MoS$_2$, the spectrum of collective excitations is angular dependent. Furthermore, for dipolar excitons in double-layer 1T$^\prime$-MoS$_2$, the normal and superfluid concentrations have a tensor form whose components depend on the direction of exciton flow. Additionally, for the double-layer, the mean-field critical temperature of superfluidity depends on the direction of exciton flow as demonstrated in Fig.\ 10. Therefore, the influence of the anisotropy of the dispersion relation of dipolar excitons in a double layer of 1T$^\prime$-MoS$_2$ with tilted Dirac bands has been investigated. We conclude that the anisotropy of the energy band structure of the
in 1T$^\prime$-MoS$_2$ exhibits superfluidity at low temperatures due to the dipole-dipole repulsion between the dipolar excitons. It is crucial to note that the binding energy of dipolar excitons and mean-field critical temperature for superfluidity are sensitive to the electron and hole effective masses. In terms of experimental observation, we can exploit some features of the photoluminescence spectrum, including emission traces caused by phonon-assisted recombination of momentum-space dark excitons. The microscopic theory for this was recently developed \cite{Malic}. The theory can be applied specifically to the case of momentum-space dark excitons in a double layer of 1T$^\prime$-MoS$_2$. This warrants further development of the formalism in conjunction with analysis of experimental results of phonon-assisted photoluminiscence experiments.
 Experimental findings include the observation of intervalley momentum-forbidden excitons influenced by compressive strain,
 serving as an ultrasensitive optical strain sensing mechanism, and the repulsion-driven propagation of dark spin-forbidden excitons,
 allowing for the propagation of valley and spin information across TMD samples for diverse optoelectronic applications \cite{GrossPoster, OpticalReadout,StrainExciton}.
 We hope that our  analytical and numerical results will provide motivation for future experimental and theoretical investigations regarding the effects of circularly polarized light on excitonic BEC and superfluidity for double layer 1T$^\prime$-MoS$_2$.

\appendix
{\section{The Hamiltonian of the charge carriers in {a monolayer of} 1T$^\prime$-MoS$_2$}
\numberwithin{equation}{section} }

The low-energy $\mathbf {k}\cdot {\mathbf p}$ Hamiltonian is
constructed based on the symmetry properties of the valence and
conduction bands of the MoS$_2$ system. The Hamiltonian of the
charge carriers in tilted a monolayer of 1T$^\prime$-MoS$_2$
{is presented in Appendix~A.}  The valence and
conduction bands mainly consist of $d$-orbitals of Mo atoms and by
$p_y$-orbitals of S atoms, respectively. $\lambda = \pm$ is used to
distinguish the locations of two independent Dirac points.
$\upsilon_1 = 3.87 \times 10^5$ m/s and $\upsilon_2 = 0.46 \times
10^5$ m/s denote the Fermi velocities along the $x$ and $y$
directions, respectively. $\upsilon_{-} = 2.86 \times 10^5$~m/s and $\upsilon_{+} = 7.21 \times 10^5$~m/s are the velocity correction
terms around the two Dirac points. The $4 \times 4$ unit matrix
${\mathbf{ I}} = \tau_0 \otimes \sigma_0$ and Dirac matrices
$\gamma_0 = \tau_1 \otimes \sigma_1$,  $\gamma_1 = \tau_2 \otimes
\sigma_0$,  $\gamma_2 = \tau_3 \otimes \sigma_0$ are defined based
on the pseudospin space $\tau_{0, 1, 2, 3}$ and Pauli matrices
$\sigma_{0, 1, 2, 3}$. In addition, ${\bf k} = (k_x, k_y)$ is the
wave vector, $\Delta = 0.042$~eV is the SOC gap, $\alpha =
|E_z/E_c|$ is the ratio of the vertical electric field over its
critical value. Taking into account the aforementioned notations,
the low-energy Hamiltonian for a 2D anisotropic tilted Dirac system
representing 1T$^\prime$~-MoS$_2$ in the vicinity of two independent
Dirac points located at (0, $\lambda\Lambda$) is given by
\cite{first}

\begin{eqnarray}
\hat{{\cal H}}_\lambda(k_x,k_y) = \hbar k_x v_1\gamma_1 + \hbar k_y
( v_2\gamma_0 -\lambda v_- {\bf I} -\lambda v_+\gamma_2) + \Delta
(\lambda \gamma_0-i\alpha\gamma_1\gamma_2)  .
\label{eq1}
\end{eqnarray}
\noindent
The energy spectrum of charge carriers in $\mathrm{1T'-MoS_{2}}$
with tilted Dirac bands in the long wavelength limit near the
extremum of the band structure  for applied electric field not close
to its critical value, $\alpha \neq 1$, is given by~\cite{BGR}
\begin{eqnarray}
  \epsilon_{\xi,s}^{\lambda}({\bf k})=  \xi |\lambda - s \alpha| \Delta+\left[  -\lambda \hbar v_{-} +\xi \hbar v_{2}\  \text{sgn}(\lambda-s\alpha) \right]k_{y}+\left(\dfrac{\xi \hbar^2 }{2\Delta|\lambda-\alpha s|}\right)\left(v_1^2 k_{x}^2+ v_{+}^2{k_y}^2\right),
\label{longwave}
\end{eqnarray}
where $\xi=\pm 1$ for the conduction (valence) band, and $s=\pm 1$
is the spin up (down) index, and we see that it depends on both
$k_y$ (a linear term) and $k_y^2$ but only on a $k_x^2$ quadratic
term. Eq.~(\ref{longwave}) also shows  that the spin-orbit coupling
opens up a gap between spin-subbands  and between the valence and
conduction bands within a chosen valley. We emphasize that
Eq.~(\ref{longwave}) is not valid in the gapless case.

\medskip
\par
\section{Electron-hole interaction in a $\mathrm{1T'-MoS_{2}}$ double layer}

\medskip
\par

The potential energy of the electron-hole attraction in this system is described by the Coulomb potential \cite{Reichman}, $V(r)=-ke^2/(\epsilon_d r_{eh})$. Making use of $r_{eh} = \sqrt{r^2 + D^2}$, where $r_{eh}$ is the distance between the electron and hole located in different parallel planes, and assuming $r\ll D$, one can expand the potential as a Taylor series in terms of $(r/D)^2$. By limiting ourselves to the first order with respect to $(r/D)^2$, we obtain
\begin{eqnarray}
V(r) = -V_0 + \gamma r^2
\label{expand}
\end{eqnarray}
\noindent
Assuming $r\ll D$ and retaining only the first two terms of the Taylor series, one
obtains the following expressions for $V_{0}$ and $\gamma$:

\begin{eqnarray}
V_{0}=\frac{ke^{2}}{\epsilon _{d}D},\hspace{1cm}\gamma =\frac{ke^{2}}{%
2\epsilon _{d}D^{3}}. \label{V0g}
\end{eqnarray}
\noindent
Replacement of $V\left(\sqrt{r^{2}+D^{2}}\right)$ by the potential in Eq.
(\ref{expand}) allows to reduce the problem of indirect exciton
formed between two layers to an exactly solvable two-body problem as
this is demonstrated in Section II.A.

\section*{Acknowledgements}

{The authors are grateful to {O.~V. Roslyak  for
valuable discussions.  The authors are grateful for support by
grants:}  GG acknowledges the support from the US AFRL Grant No.
FA9453-21- 1-0046.

\end{document}